\documentclass[submission,Phys]{SciPost}
\usepackage{amsmath,amssymb,mathtools,mathrsfs,bm}
\usepackage{empheq}
\usepackage{amsthm}
\usepackage{booktabs,array}
\usepackage{graphicx}
\usepackage{tikz}
\usetikzlibrary{arrows.meta,calc,patterns,positioning}
\usepackage{xcolor}
\usepackage{microtype}
\hypersetup{colorlinks=true,linkcolor=scipostdeepblue,citecolor=scipostdeepblue,urlcolor=scipostdeepblue}

\definecolor{deepblue}{RGB}{30,70,105}
\definecolor{teal}{RGB}{15,125,125}
\definecolor{pale}{RGB}{233,243,247}
\definecolor{gold}{RGB}{176,123,20}

\newcommand{\dr}{\mathrm{dr}}

\newcommand{\supp}{\operatorname{supp}}
\newcommand{\gpol}{{\gamma_{\mathrm {pol}}}}
\newcommand{\pol}{\mathrm{pol}}

\theoremstyle{definition}

\theoremstyle{remark}

\begin{document}

\begin{center}
{\Large\bfseries Emptiness formation in quantum fluids: \\ a hydrodynamic solution}
\end{center}

\begin{center}
Alexander G. Abanov\textsuperscript{1,*} and Boris A. Khanikati\textsuperscript{2}
\end{center}

\begin{center}
\textbf{1} Department of Physics and Astronomy, Stony Brook University,\\
Stony Brook, NY 11794, USA\\[0.3em]
\textbf{2} Department of Physics, Harvard University, Cambridge, MA 02138, USA
\end{center}

\begin{center}
\textsuperscript{*}\href{mailto:alexandre.abanov@stonybrook.edu}{alexandre.abanov@stonybrook.edu}
\end{center}
\begin{center}
\today
\end{center}


\section*{Abstract}
{\bfseries\boldmath
We develop an auxiliary-kernel method for constructing the hydrodynamic instantons that govern macroscopic emptiness formation in one-dimensional Galilean quantum fluids.
The key step is to represent the equation of state by an auxiliary integral kernel inspired by the Bethe ansatz. For equations of state admitting such a representation, we reduce the imaginary-time hydrodynamic problem to a single integral equation. Its solution determines the full density and current profiles, the vacuum boundary, and the leading emptiness rate. Notably, the construction does not require microscopic integrability. For the Lieb--Liniger gas, we derive the previously conjectured integral equation for the leading rate and use a known elliptic solution to obtain explicit expressions for the rate and vacuum boundary. Direct hydrodynamic numerics support the predicted boundary. We also recover known instanton solutions for a family of polytropic fluids.

\par}

\vspace{10pt}
\noindent\rule{\textwidth}{1pt}
{\small\setlength{\cftbeforesecskip}{0.5em}\tableofcontents}\thispagestyle{fancy}
\noindent\rule{\textwidth}{1pt}
\vspace{10pt}
\clearpage

\section{Introduction}

The \textit{emptiness formation probability} (EFP) probes rare, nonlocal density fluctuations in one-dimensional quantum fluids. For a ground state with uniform density $\rho_0$, the EFP is the probability $P(R)$ that an interval of length $2R$ contains no particles. This quantity arises naturally in the study of integrable models, where integrability techniques often make exact and asymptotic results tractable. It therefore provides a useful benchmark for nonperturbative methods \cite{kitanine2002xxz,kitanine2002XXZ-asymptotic,korepin2003xxz,colomo2008emptiness,abanov2002probability,AbanovFranchini2003}.

When $R$ is large compared to microscopic lengths, such as the average interparticle distance, emptiness requires a collective rearrangement of the fluid. The leading probability exponent is then governed semiclassically by an \textit{emptiness instanton}, a solution of the Euler equations in imaginary time that features an empty interval at $\tau=0$ and approaches the uniform state at spacetime infinity \cite{abanov-hydro}. The action of this saddle determines the cost of the fluctuation. The empty region persists over a finite interval of imaginary time and is bounded by a \textit{vacuum boundary}\footnote{Such boundaries are also called \emph{arctic curves}, by analogy with the boundaries between frozen and fluctuating regions in random tilings~\cite{StephanRandomTilings}.}, whose astroid-like geometry has been studied in hydrodynamic treatments of the EFP \cite{abanov-hydro,yeh2022emptiness,abanov2025polytropicEFP}. 
This connects emptiness formation to limit-shape phenomena, in which macroscopic regions are separated by sharply defined boundaries \cite{Kenyon2007,StephanRandomTilings,de2021limit,Pallister_2022}.

For a Galilean fluid, the dispersionless instanton problem is determined entirely by the equation of state $\epsilon(\rho)$, regardless of whether the microscopic model is integrable. The challenge is to solve the nonlinear hydrodynamic equations and also determine the vacuum boundary. We use Riemann invariants and a hodograph transformation to exchange the physical coordinates with the hydrodynamic fields. This leads to a linear equation for a single potential $\mathcal W$, with coefficients determined by the local Luttinger parameter. Once $\mathcal W$ is known, the inverse transformation reconstructs the physical instanton. The problem thus reduces to solving a linear mixed boundary-value problem for $\mathcal W$, with boundary conditions fixed by the emptiness constraint and the uniform background.

To solve this mixed problem, we introduce an \emph{auxiliary-kernel method} (Sec.~\ref{sec:auxiliary-kernel}), inspired by the dressing formalism of Bethe-ansatz thermodynamics~\cite{korepin1997quantum,doyon2020lecture}. We show that an even difference kernel representing the equation of state generates the spectral modes of the linear hodograph equation through the associated dressing operation. Using these modes, we reduce the mixed boundary conditions to a single integral equation on a finite interval. Its solution determines the instanton profile, vacuum boundary, and action. Remarkably, this use of Bethe-ansatz techniques does not require microscopic integrability.

We first apply this construction to the repulsive Lieb--Liniger Bose gas, using its physical two-body scattering kernel. In previous work~\cite{khanikati2026efp}, starting from the dual-field representation of the EFP~\cite{korepin1995probabilityphaseseparationbose}, we conjectured its leading large-$R$ rate in terms of an integral equation. Here, the general auxiliary-kernel method yields this equation directly from hydrodynamics. For the Lieb--Liniger kernel, the resulting integral equation can be solved explicitly. After rescaling, its resolvent form coincides with the planar saddle-point equation of Hoppe's two-matrix model~\cite{hoppe1982thesis}, allowing us to use the elliptic solution of Kazakov, Kostov, and Nekrasov (KKN)~\cite{kazakov1999dparticles}. This gives explicit expressions for the rate and vacuum boundary. We obtain the full liquid profile by combining the elliptic solution with Lieb--Liniger thermodynamics.

Polytropic fluids illustrate how the auxiliary-kernel method extends beyond microscopic integrability. The equation of state has the form $\epsilon(\rho)\propto\rho^{\gpol}$, and their emptiness instantons are known from a direct hydrodynamic calculation~\cite{abanov2025polytropicEFP}. We construct singular auxiliary kernels for $1<\gpol\leq3$ that reproduce the equation of state and recover the known instanton, vacuum boundary, and action.

The paper is organized as follows. Section~\ref{sec:problem} defines the observables and specifies the macroscopic limit. Section~\ref{sec:main-results} summarizes our main results. The hydrodynamic formulation, auxiliary-kernel method, and applications follow in Secs.~\ref{sec:hydrodynamics}--\ref{sec:polytropic}.
The appendices provide the detailed derivations, endpoint estimates, and numerical methods.

\section{Statement of the problem and macroscopic limit}\label{sec:problem}

\subsection{Emptiness probability and conditioned density profile}

Let $|\Psi_0\rangle$ be a translation-invariant ground state of a one-dimensional quantum fluid with density $\rho_0$. The microscopic density operator is $\widehat\rho(x)=\sum_i\delta(x-\widehat x_i)$, where $\widehat x_i$ is the position operator of particle $i$. The particle number in $[-R,R]$ is
\begin{equation}
    \widehat N_R=\int_{-R}^{R}\widehat\rho(x)\,dx.
\end{equation}
The projector $\widehat\Pi_R$ onto the zero-particle sector defines the EFP as
\begin{equation}\label{eq:efp-definition}
    P(R)=\langle\Psi_0|\widehat\Pi_R|\Psi_0\rangle,
    \qquad
    \widehat\Pi_R=\lim_{\alpha\to+\infty}e^{-\alpha\widehat N_R}.
\end{equation}
Equivalently, the EFP is the limit of the \textit{full counting statistics} generating function, $P(R)=\lim_{\alpha\to+\infty}Z_R(\alpha)$, where
\begin{equation}
    Z_R(\alpha)=\langle\Psi_0|e^{-\alpha\widehat N_R}|\Psi_0\rangle
    =\sum_{n=0}^{\infty}p_R(n)e^{-\alpha n},
    \qquad p_R(n)=\Pr(N_R=n).
\end{equation}
The mean particle number in the interval is $2\rho_0R$, while the EFP selects $N_R=0$.

In units $\hbar=1$, the imaginary-time density operator is
\begin{equation}
    \widehat\rho(x,\tau)=e^{\tau \widehat H}\widehat\rho(x)e^{-\tau \widehat H}.
\end{equation}
Correspondingly, the density profile conditioned on an empty interval at $\tau=0$ is
\begin{equation}\label{eq:conditioned-density}
    \rho_{\mathrm{EFP}}(x,\tau)=
    \frac{\langle\Psi_0|\mathcal T_\tau
    [\widehat\rho(x,\tau)\widehat\Pi_R]|\Psi_0\rangle}{P(R)},
\end{equation}
where $\mathcal T_\tau$ denotes imaginary-time ordering. At $\tau=0$, this reduces to the conditional density for a position measurement, with
\begin{equation}
    \rho_{\mathrm{EFP}}(x,0)=0,\qquad |x|<R.
\end{equation}
At nonzero $\tau$, Eq.~\eqref{eq:conditioned-density} describes an imaginary-time correlation conditioned on the empty interval.

\subsection{Macroscopic limit}\label{sec:macroscopic-limit}
We take $R\to\infty$ at fixed background state and interaction parameters, with $R$ large compared to microscopic length scales and $\rho_0R\gg1$. The dimensionless EFP rate $f$, which depends on the fluid and its background state, is defined by
\begin{equation}\label{eq:rate-definition}
    -\log P(R)=(\rho_0R)^2 f+o(R^2).
\end{equation}

We compute this rate, the density and current fields, and the vacuum boundary using dispersionless hydrodynamics for a Galilean fluid. The instanton has spatial and imaginary-time scales $R$ and $R/c_{s0}$, where $c_{s0}$ is the background sound speed. In the scaled coordinates $\xi=x/R$ and $\theta=c_{s0}\tau/R$, the dispersionless equations and EFP boundary conditions are independent of $R$. The action therefore has the $R^2$ scaling in Eq.~\eqref{eq:rate-definition}~\cite{abanov-hydro}, provided the action integral converges. This requires particular attention near the endpoints $(x,\tau)=(\pm R,0)$ of the empty interval, where the density is expected to become singular.\footnote{Appendix~\ref{app:endpoint-estimates} gives sufficient conditions for endpoint convergence and estimates the hydrodynamic contribution from small neighborhoods of the endpoints.} The corresponding macroscopic density profile is defined by
\begin{equation}\label{eq:macroscopic-profile}
    \frac{\rho_{\mathrm{EFP}}(R\xi,R\theta/c_{s0})}{\rho_0}
    \longrightarrow \varrho(\xi,\theta).
\end{equation}
The convergence is understood for the density coarse-grained over microscopic scales, away from microscopic boundary layers. The profile approaches unity at spacetime infinity and vanishes at $\theta=0$ for $|\xi|<1$.

We assume that the hydrodynamic instanton defines single-valued density and current fields throughout the liquid region and is the minimizing saddle under the emptiness constraint. Identifying its action with the leading EFP exponent further assumes that gradient corrections and microscopic endpoint regions contribute only at subleading order. These assumptions apply throughout the paper and are not guaranteed by solving the hydrodynamic equations. The method in Sec.~\ref{sec:auxiliary-kernel} also requires a kernel representation of the equation of state, together with the mathematical assumptions stated there.

\subsection{The Lieb--Liniger gas}\label{sec:problem-ll}
Our primary example is the repulsive Lieb--Liniger Bose gas. In units $m = \hbar = 1$, the $N$-particle Hamiltonian is
\begin{equation}
    \widehat H=-\frac12\sum_{j=1}^{N}\partial_{x_j}^2
    +c\sum_{i<j}\delta(x_i-x_j),\qquad c>0.
\end{equation}
Here, $c$ is the contact interaction strength. In the thermodynamic limit at ground-state density $\rho_0$, the dimensionless coupling is $\gamma_0=c/\rho_0$, and dimensional analysis implies that the rate depends only on it, $f=f(\gamma_0)$.

To obtain the equation of state, we consider ground states at fixed $c$ and varying density. In the rest frame, the occupied rapidities fill an interval $[-Q,Q]$, whose endpoint $Q$ parametrizes these states. 
The ground-state rapidity density $\rho_p(\lambda;Q)$ satisfies the Lieb integral equation~\cite{lieb1963exact,korepin1997quantum},
\begin{equation}\label{eq:problem-ll-thermodynamics}
    \rho_p(\lambda;Q)=\frac{1}{2\pi}+\int_{-Q}^{Q}\frac{d\mu}{2\pi}
    \Delta(\lambda-\mu)\,\rho_p(\mu;Q),
    \qquad |\lambda|\leq Q,
\end{equation}
where $\Delta(\lambda-\mu)$ is the two-particle scattering kernel, given by
\begin{equation}\label{eq:ll-scattering-kernel}
    \Delta(\lambda)=\frac{2c}{c^2+\lambda^2}.
\end{equation}
The particle density and energy per unit length are then
\begin{equation}
    \rho(Q)=\int_{-Q}^{Q}\rho_p(\lambda;Q)\,d\lambda,
    \qquad
    \epsilon(Q)=\int_{-Q}^{Q}\frac{\lambda^2}{2}\,\rho_p(\lambda;Q)\,d\lambda.
\end{equation}
Eliminating $Q$ between these two functions gives the equation of state $\epsilon(\rho)$ at fixed $c$. In the hydrodynamic description, $Q$ is the \textit{local Fermi rapidity}, measured in the local rest frame. Its value $q$ in the uniform background is fixed by $\rho(q)=\rho_0$ and is called the \textit{background Fermi rapidity}.

\section{Main results}\label{sec:main-results}
\paragraph*{General construction.} Following the hydrodynamic approach to emptiness formation~\cite{abanov-hydro}, we formulate the EFP as an instanton problem in Sec.~\ref{sec:hydrodynamics}. The instanton produces an empty interval $[-R,R]$ at $\tau=0$ and approaches the uniform fluid at spacetime infinity. We call this imposed interval in spacetime the \emph{slit}; the vacuum region itself extends to nonzero imaginary times. Using the density and velocity as coordinates reduces the Euler equations to a linear mixed boundary-value problem for a single potential (Fig.~\ref{fig:boundary_conditions}).

In Sec.~\ref{sec:auxiliary-kernel}, we draw on Bethe-ansatz techniques to solve this problem. An auxiliary integral kernel representation of the equation of state allows us to construct the spectral modes~\eqref{eq:dressed-spectral-mode} of the linear equation through a dressing operation. The mixed boundary conditions then reduce to a single integral equation on a finite interval. Given a regular solution, we obtain analytic expressions for the instanton density and current, the vacuum boundary~\eqref{eq:general-vacuum-boundary}, and the action that gives the leading EFP rate~\eqref{eq:main-general-response}. In Sec.~\ref{sec:polytropic}, we verify the construction separately for a family of polytropic gases, demonstrating the applicability of the method beyond microscopic integrability.

We state the Lieb--Liniger results below and derive them in Sec.~\ref{sec:ll-application}. Under the physical assumptions stated in Sec.~\ref{sec:macroscopic-limit}, they yield the leading large-$R$ EFP exponent and the macroscopic conditioned density.

\subsection{Rate function}
For the Lieb--Liniger gas, we normalize the rate by the background Fermi rapidity $q$, using the dimensionless ratio $r=c/q$ as the coupling:
\begin{equation}\label{eq:rescaled-rate-definition}
    -\log P(R)=(qR)^2\widehat f(r)+o(R^2).
\end{equation}
The rate in the density normalization of Eq.~\eqref{eq:rate-definition} is recovered through $f(\gamma_0)=(q/\rho_0)^2\widehat f(r)$.

At fixed $r$, write the rapidity as $\lambda=qt$, with the odd dimensionless function $g(t)$ on $[-1,1]$ defined by the integral equation
\begin{equation}\label{eq:main-ll-response}
    \int_{-1}^{1}g(s)\log\left(1+\frac{r^2}{(t-s)^2}\right)ds=4t,
    \qquad -1<t<1.
\end{equation}
The rescaled EFP rate is then determined by its first moment:
\begin{equation}\label{eq:main-ll-rate}
    \widehat f(r)=\frac12\int_{-1}^{1}t\,g(t)\,dt.
\end{equation}
This hydrodynamic rate agrees with the leading EFP asymptotics conjectured in Ref.~\cite{khanikati2026efp}, written here in the $(qR)^2$ normalization. Equation~\eqref{eq:main-ll-response} can be solved numerically. Alternatively, we can rewrite it as a resolvent equation and use KKN's elliptic solution. This gives explicit expressions for $g$, the rate, and the vacuum boundary (Sec.~\ref{sec:ll-elliptic-rate}).

\begin{figure}[htbp]
    \centering
    \includegraphics[width=0.94\linewidth]{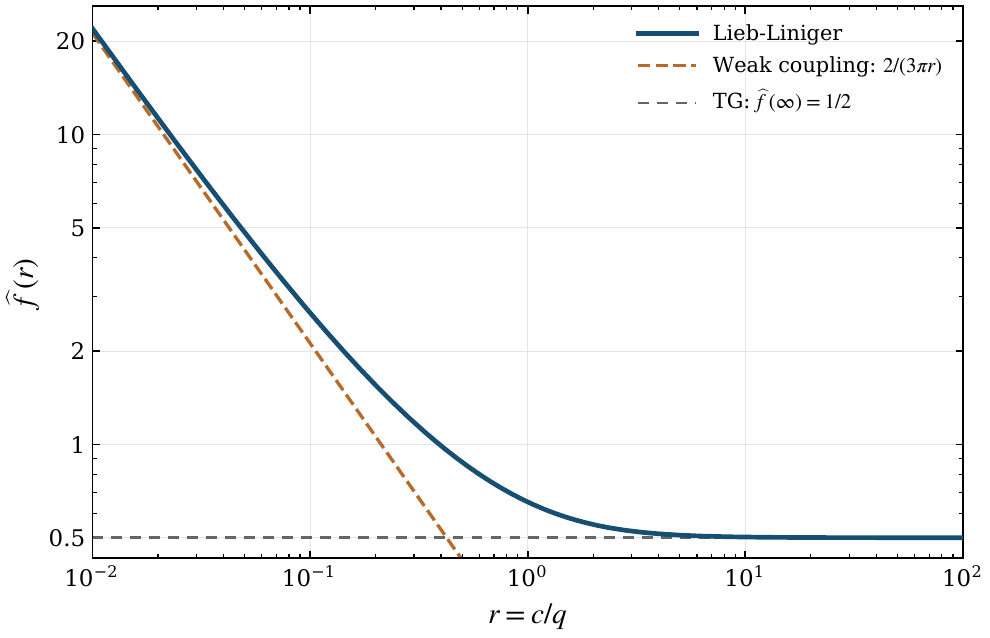}
    \caption{Rescaled Lieb--Liniger EFP rate $\widehat f(r)$ versus $r=c/q$, with $q$ the background Fermi rapidity. The solid curve is obtained from the integral response equation~\eqref{eq:main-ll-response} and the moment~\eqref{eq:main-ll-rate}; both axes are logarithmic. The dashed curves show the leading weak-coupling asymptote $2/(3\pi r)$ and the Tonks--Girardeau limit $1/2$. The leading EFP exponent is $(qR)^2\widehat f(r)$.}
    \label{fig:ll-efp-rate}
\end{figure}

The Lieb equation~\eqref{eq:problem-ll-thermodynamics} determines $q/\rho_0$ and relates $r$ to the physical coupling through $\gamma_0=rq/\rho_0$. Figure~\ref{fig:ll-efp-rate} shows $\widehat f(r)$ from weak to strong repulsion.

\subsection{Vacuum boundary}
The same function $g$ determines the vacuum boundary. On its upper-right branch ($x>0$, $\tau>0$), let $v\leq0$ be the imaginary-time fluid velocity. Introducing $\eta=|v|/q$, define
\begin{equation}\label{eq:main-ll-b}
    b(\eta)=1-\int_0^1t\,g(t)
    \left[\frac{1}{t^2+\eta^2}-\frac{1}{t^2+(\eta+r)^2}\right]dt.
\end{equation}
The upper-right branch of the vacuum boundary is then given parametrically by
\begin{equation}\label{eq:main-ll-boundary}
    \frac{x(\eta)}{R}=b(\eta)-\eta b'(\eta),\qquad
    \frac{q\tau(\eta)}{R}=b'(\eta),\qquad \eta\geq0,
\end{equation}
with endpoint values understood as limits. As $\eta$ increases, the arc runs from the closing point on the imaginary-time axis to the slit endpoint $(R,0)$. Reflections in $x$ and $\tau$ complete the curve, enclosing the vacuum region.

\begin{figure}[!t]
    \centering
    \includegraphics[width=\linewidth]{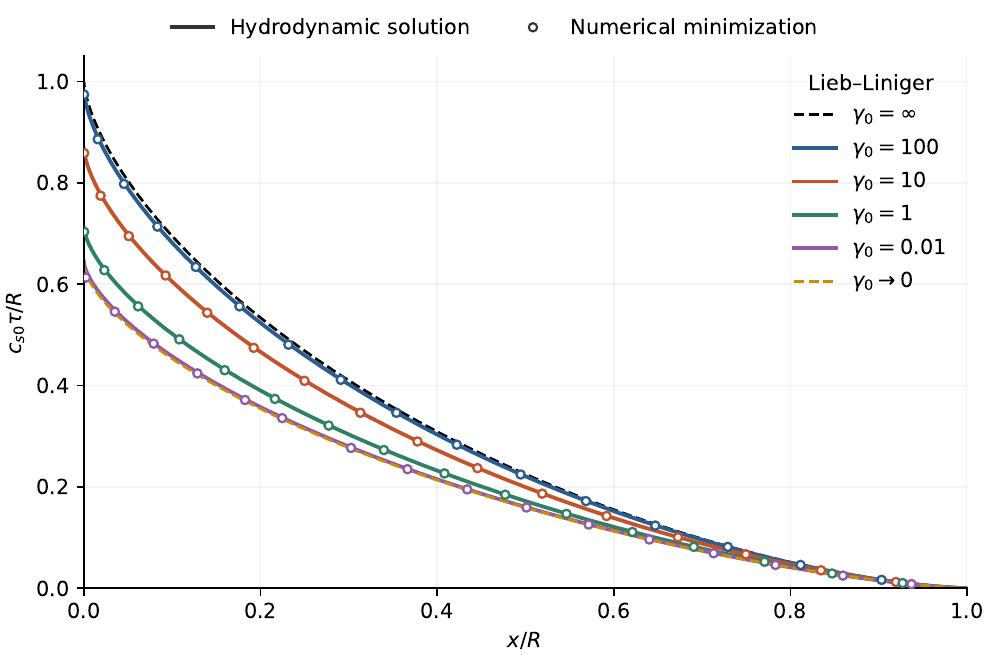}
    \caption{Lieb--Liniger vacuum boundaries for $\gamma_0=100,10,1,0.01$, shown in coordinates $x/R$ and $c_{s0}\tau/R$. Dashed curves show the exact limits $\gamma_0=\infty$ and $\gamma_0\to0$, while solid curves show the hydrodynamic prediction~\eqref{eq:main-ll-boundary}. Open circles show numerical results obtained by extrapolation from the liquid side, with matching colors. Numerical points are omitted in unresolved regions. See Sec.~\ref{sec:hydro-numerics} and Appendix~\ref{app:hydro-numerics} for details.}
    \label{fig:vacuum-boundary-comparison}
\end{figure}

Direct minimization of the Euclidean hydrodynamic action supports the predicted vacuum boundary (Fig.~\ref{fig:vacuum-boundary-comparison}). The calculation uses the analytic leading far-field amplitude, but determines the vacuum boundary and closing time numerically. Section~\ref{sec:hydro-numerics} discusses the accuracy and convergence limitations, with implementation details in Appendix~\ref{app:hydro-numerics}.

\subsection{Full instanton}
Our construction yields not only the vacuum boundary but the full hydrodynamic instanton, that is, the density $\rho(x,\tau)$ and current $j(x,\tau)=\rho(x,\tau)v(x,\tau)$ throughout spacetime. These fields describe how the liquid is redistributed as the empty region forms and closes in imaginary time, and how the configuration approaches the uniform fluid far from the fluctuation.

The same function $g$ that determines the rate and vacuum boundary also determines these fields when combined with the Lieb--Liniger thermodynamics. The general reconstruction procedure is summarized in Sec.~\ref{sec:practical-procedure}. For the Lieb--Liniger gas, Eq.~\eqref{eq:ll-profile-quadrature} and the coordinate map~\eqref{eq:hodograph-mapping} give the density and current fields, as explained in Sec.~\ref{sec:ll-reconstruction}. The resulting density gives the macroscopic conditioned profile~\eqref{eq:macroscopic-profile}. Figure~\ref{fig:instanton-density-profiles} shows density profiles obtained by numerical minimization of the hydrodynamic action.

\section{Hydrodynamic formulation}\label{sec:hydrodynamics}
\subsection{Action and hydrodynamic instanton}\label{sec:hydro-action}
Consider a one-dimensional Galilean fluid with energy density $\epsilon(\rho)$, in units $m=\hbar=1$ and with $\epsilon(0)=0$. For density $\rho$ and imaginary-time current $j$, the Euclidean action is
\begin{equation}
    \mathcal I(\rho,j) = \int dx\,d\tau\, \left[\frac{j^2}{2\rho} + U(\rho) \right], \quad U(\rho) = \epsilon(\rho)-\epsilon(\rho_0) -\mu_0(\rho-\rho_0),
\end{equation}
where $\rho_0$ is the background density, $\mu(\rho)=\epsilon'(\rho)$, and $\mu_0=\mu(\rho_0)$. The background energy and chemical potential terms are subtracted so that the uniform background at rest contributes zero action. Write $j=\rho v$, where $v$ is the imaginary-time fluid velocity. Varying the action subject to continuity gives
\begin{align}
    \partial_\tau \rho + \partial_x(\rho v) = 0, \\
    \partial_\tau v + v\partial_x v = \mu'(\rho)\partial_x\rho.
\end{align}
We assume a positive sound-speed squared, $c_s^2=\rho\mu'(\rho)>0$, in the liquid. The pressure term has the opposite sign from real-time dynamics, making this system elliptic. The EFP instanton is the saddle with an empty interval at $\tau=0$ that approaches the uniform background at spacetime infinity. Under the physical assumptions of Sec.~\ref{sec:macroscopic-limit}, its action determines the leading EFP exponent,
\begin{equation}
    - \log P(R) = \mathcal I_{\rm EFP} + o(R^2), \qquad \mathcal I_{\rm EFP} = O(R^2).
\end{equation}

\subsection{Hodograph potential and the EFP mixed boundary-value problem}\label{sec:hodograph-problem}

To rewrite the Euler equations in terms of the local fluid state, introduce the acoustic density coordinate $Q$ and its background value $q$:
\begin{equation}\label{eq:acoustic-coord}
    Q(\rho) = \int_0^\rho \frac{c_s(\rho')}{\rho'}\,d\rho', \quad q = Q(\rho_0).
\end{equation}
For the Lieb--Liniger gas, $q$ is the background Fermi rapidity introduced in Sec.~\ref{sec:problem-ll}, as shown by the thermodynamic identities in Sec.~\ref{sec:auxiliary-kernel}. By reflection symmetries in space and imaginary time, it suffices to solve the problem in the upper-right physical quadrant $x,\tau>0$, where $v \leq 0$. It is convenient to introduce the Riemann invariants, which diagonalize the Euler equations \cite{landauFluidMech},
\begin{equation}
    r_\pm = -iv \pm Q, \qquad  \partial_\tau r_{\pm} + (v \pm ic_s) \partial_x r_\pm = 0. 
\end{equation}
The characteristic velocities depend on both invariants, so diagonalization alone does not decouple the problem. The thermodynamic weight entering the hodograph transformation is
\begin{equation}
    K(\rho) = \pi \frac{d\rho}{dQ} = \frac{\pi \rho}{c_s}.
\end{equation}
For a one-component Galilean quantum fluid, $K$ is the local Luttinger parameter. A real-time sound mode satisfies $\delta\rho=\pm(K/\pi)\delta u$, where $u$ is the physical real-time velocity. Below, $K(Q)$ denotes the same quantity expressed in terms of the acoustic coordinate.

A hodograph transformation makes the fluid state $(Q, v)$ the independent variables. Where the map from $(x,\tau)$ to $(Q,v)$ is locally invertible, write the physical coordinates as $x(Q,v)$ and $\tau(Q,v)$. The hydrodynamic equations then take the form
\begin{equation}\label{eq:inverse-hydrodynamics}
    x_Q=v\tau_Q+c_s\tau_v,\qquad x_v=v\tau_v-c_s\tau_Q.
\end{equation}
Since $c_s$ depends only on $Q$, the first equation reads $\partial_v(c_s\tau)=\partial_Q(x-R-v\tau)$. This compatibility condition ensures the local existence of a potential $\mathcal W$, defined up to an additive constant by
\begin{equation}
    \mathcal W_Q = c_s \tau, \qquad \mathcal W_v = x - R - v\tau.
\end{equation} 
Substituting these definitions into the second equation and using $\partial_Q(Kc_s)=K$ gives the linear hodograph equation
\begin{equation}\label{eq:hodograph-equation}
    \partial_Q(K\mathcal W_Q) + \partial_v (K \mathcal W_v) = 0.
\end{equation}
Once $\mathcal W$ is known, the inverse map is
\begin{equation}\label{eq:hodograph-mapping}
    \tau = \frac{\mathcal W_Q}{c_s}, \qquad x = R + \mathcal W_v + v\tau.
\end{equation}
Together with $\rho=\rho(Q)$, this reconstructs the liquid configuration. We assume that the resulting map is globally one-to-one, as discussed in Sec.~\ref{sec:macroscopic-limit}, since this does not follow from local invertibility or the linear equation alone.

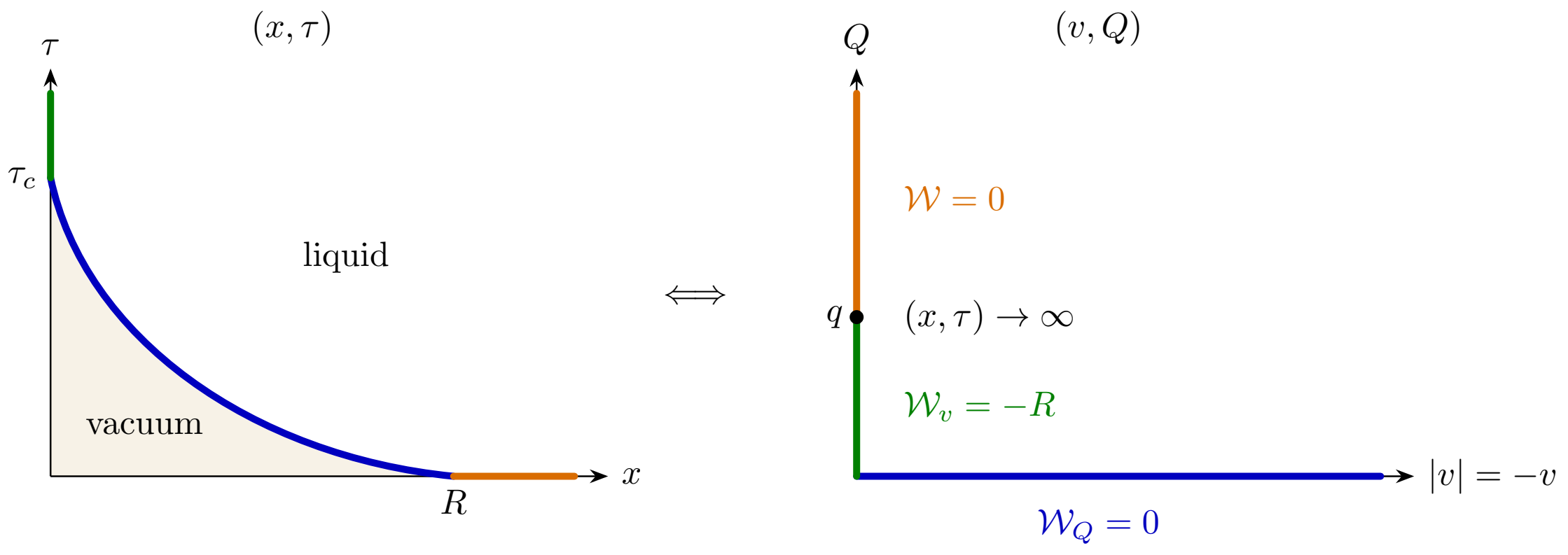
\begin{figure}[t!]
    \centering
    \resizebox{\linewidth}{!}{%
    \begin{tikzpicture}[>=Stealth,x=1.35cm,y=1cm,font=\small,
        boundary/.style={line width=2pt,line cap=round},
        vacuumedge/.style={boundary,draw=blue!75!black},
        symmetryedge/.style={boundary,draw=green!50!black},
        exterioredge/.style={boundary,draw=orange!85!black}]
    \begin{scope}
    \fill[gold!10] (0,0)--(0,3) .. controls (0.25,1.4) and (1.5,0.2) .. (3,0)--cycle;
    \draw[->,line width=0.5pt] (0,0)--(4.15,0) node[right] {$x$};
    \draw[->,line width=0.5pt] (0,0)--(0,4.1) node[above] {$\tau$};
    \draw[vacuumedge] (0,3) .. controls (0.25,1.4) and (1.5,0.2) .. (3,0);
    \draw[symmetryedge] (0,3)--(0,3.85);
    \draw[exterioredge] (3,0)--(3.9,0);
    \node[left] at (0,3) {$\tau_c$}; \node[below] at (3,0) {$R$};
    \node at (0.7,0.5) {vacuum}; \node at (2.2,2.2) {liquid};
    \node at (1.8,4.5) {$(x,\tau)$};
    \end{scope}
    \node at (4.8,1.8) {$\Longleftrightarrow$};
    \begin{scope}[shift={(6,0)}]
    \draw[->,line width=0.5pt] (0,0)--(4.15,0) node[right] {$|v|=-v$};
    \draw[->,line width=0.5pt] (0,0)--(0,4.1) node[above] {$Q$};
    \draw[vacuumedge] (0,0)--(3.9,0);
    \draw[symmetryedge] (0,0)--(0,1.6);
    \draw[exterioredge] (0,1.6)--(0,3.85);
    \fill (0,1.6) circle (2pt) node[left] {$q$};
    \node[right] at (0.25,1.6) {$(x,\tau)\to\infty$};
    \node[right,text=orange!85!black] at (0.25,2.8) {$\mathcal W=0$};
    \node[right,text=green!50!black] at (0.25,0.7) {$\mathcal W_v=-R$};
    \node[below,text=blue!75!black] at (1.8,-0.2) {$\mathcal W_Q=0$};
    \node at (1.8,4.5) {$(v,Q)$};
    \end{scope}
    \end{tikzpicture}
    }
    \caption{Mapping of the liquid region in the upper-right physical quadrant to the hodograph quadrant. Matching colors identify the vacuum boundary, mapped to $Q=0$ (blue); the liquid part of $x=0$, $\tau>\tau_c$, mapped to $v=0$, $0<Q<q$ (green); and the slit exterior $\tau=0$, $x>R$, mapped to $v=0$, $Q>q$ (orange). Here, $\tau_c$ is the closing time of the vacuum region.}
    \label{fig:boundary_conditions}
\end{figure}

To determine $\mathcal W$, we now translate the physical EFP conditions into boundary data. The imposed empty interval, $\rho(x,0)=0$ for $|x|<R$, lies within the vacuum region. Figure~\ref{fig:boundary_conditions} shows how the surrounding liquid maps to the fixed hodograph quadrant $Q\geq0$, $v\leq0$. The liquid portions of the physical axes map to $v=0$, while the vacuum boundary maps to $Q=0$. The unknown shape of this curve will be recovered from the inverse map.

Reflection symmetry gives $v=0$ on both liquid portions of the physical axes, but they correspond to different ranges of $Q$. The slit exterior, $\tau=0$, $x>R$, maps to $Q>q$. Since $\tau=0$, the defining relation $\mathcal W_Q=c_s\tau$ gives $\mathcal W_Q=0$ on this segment. The potential is therefore constant there, and we choose its additive constant so that $\mathcal W=0$.

The liquid part of the spatial symmetry axis, $x=0$, $\tau>\tau_c$, maps to $0<Q<q$, where $\tau_c$ is the closing time of the vacuum region. Setting $x=v=0$ in the defining relation for $\mathcal W_v$ gives $\mathcal W_v=-R$. Thus the same hodograph boundary $v=0$ carries a Dirichlet condition for $Q>q$ and a Neumann condition for $0<Q<q$.

We approach the vacuum boundary $Q=0$ from the liquid, taking $Q\to0$ at fixed $v<0$. We assume that $c_s\to0$ and require the imaginary time $\tau=\mathcal W_Q/c_s$ to remain finite, from which it follows that $\mathcal W_Q\to0$.

Physical spacetime infinity, where $\rho\to\rho_0$ and $v\to0$, maps to the junction $(Q,v)=(q,0)$. The slit endpoint $(x,\tau)=(R,0)$ is reached at hodograph infinity, $Q^2+v^2\to\infty$, where we require the potential to decay. Collecting the boundary conditions gives (see Figure~\ref{fig:boundary_conditions})
\begin{align}
    \mathcal W(Q,0) &= 0, && Q>q, \label{eq:hodograph-dirichlet}\\
    \mathcal W_v(Q,0) &= -R, && 0<Q<q, \label{eq:hodograph-neumann}\\
    \mathcal W_Q(0,v) &= 0, && v<0, \\
    \mathcal W(Q,v) &\to 0, && Q^2+v^2\to\infty.
\end{align}
In the same liquid-side limit, $K\mathcal W_Q=\pi\rho\tau\to0$ as $Q\to0$, since $\rho\to0$ and $\tau$ remains finite. For singular $K$, this weighted condition is needed to specify the spectral operator domain, since $\mathcal W_Q\to0$ alone does not imply it.

\subsection{Spectral representation}\label{sec:spectral}
Because $K$ depends only on $Q$, the equation is translation invariant in $v$ and admits separated modes of the form $B_p(Q)e^{-p|v|}$. With $|v|=-v$ on the chosen quadrant, expand the decaying potential as
\begin{equation}\label{eq:spectral-superposition}
    \mathcal W = - \int_0^\infty a(p) B_p(Q) e^{-p|v|} \,dp.
\end{equation}
The mixed boundary conditions at $v=0$ will fix the amplitudes $a(p)$. The modes $B_p$ obey
\begin{equation}\label{eq:spectral-diffeq}
    \frac{1}{K(Q)}\partial_Q[K(Q) \partial_Q B_p] = - p^2 B_p.
\end{equation}
We impose $B_p'(0)=0$ to satisfy the vacuum boundary condition $\mathcal W_Q(0,v)=0$ for $v<0$, and normalize the modes by $B_p(0)=1$. For singular $K$, we also require $KB_p'\to0$ as $Q\to0$. The corresponding Sturm--Liouville operator is
\begin{equation}
    \mathcal L = -K^{-1} \partial_Q(K \partial_Q),
\end{equation}
with the natural inner product
\begin{equation}
    \langle f, g \rangle_K = \int_0^\infty Kfg \,dQ. 
\end{equation}

To determine the amplitudes, it is sufficient to find the potential on the boundary $v=0$. Define
\begin{equation}\label{eq:potential-trace}
    F(Q)=-\mathcal W(Q,0).
\end{equation}
The decaying solution with this boundary value is $\mathcal W(Q,v)=-e^{-|v|\sqrt{\mathcal L}}F(Q)$, with $\mathcal L$ defined using the vacuum boundary conditions stated above. Since $|v|=-v$, differentiation in $v$ multiplies each mode by $p$. Thus, the Dirichlet-to-Neumann operator $\Lambda=\sqrt{\mathcal L}$ relates the boundary values to the normal derivative through $\Lambda F=-\partial_v\mathcal W|_{v=0^-}$. The two EFP conditions become
\begin{align}
    F(Q) & = 0, \qquad Q>q, \\
    \Lambda F(Q) &= R, \qquad 0<Q<q.
\end{align}
The equation of state fixes $\mathcal L$, while solving the mixed boundary problem for $F$ determines the amplitudes $a(p)$, and hence the full potential $\mathcal W$.

\subsection{Vacuum boundary and instanton action}\label{sec:hydro-observables}
\paragraph*{Vacuum boundary.} Suppose that we have found the hodograph potential $\mathcal W(Q,v)$ satisfying the mixed boundary conditions. The inverse map~\eqref{eq:hodograph-mapping} then determines the liquid configuration. The vacuum boundary follows by taking the limit $Q \to 0$, where the density vanishes. The second relation in Eq.~\eqref{eq:hodograph-mapping} then becomes
\begin{equation}\label{eq:vacuum-characteristic}
    x-v\tau=X(v),\qquad X(v)\equiv R+\partial_v\mathcal W(0,v).
\end{equation}
For each $v<0$, this describes a limiting characteristic line tangent to the vacuum boundary, with intercept $X(v)$ at $\tau=0$. The boundary is the envelope of these lines as $v$ varies. Differentiating Eq.~\eqref{eq:vacuum-characteristic} with respect to $v$ at fixed $(x,\tau)$ gives the envelope condition $-\tau=X'(v)$, and hence
\begin{equation}\label{eq:vacuum-envelope}
    \tau(v)=-X'(v), \qquad x(v)=X(v)-vX'(v).
\end{equation}
Thus, the vacuum boundary is determined entirely by $\mathcal W(0,v)$ and does not require the full liquid configuration.

\paragraph*{Optimal action.} The optimal action can also be evaluated without reconstructing the full instanton profile. We show in Appendix~\ref{app:hodograph-action} that, on the instanton configuration,
\begin{equation}\label{eq:action-dirichlet}
    \mathcal I_{\mathrm{EFP}} = E_D,
\end{equation}
where the weighted Dirichlet energy is defined by
\begin{equation}\label{eq:dirichlet-definition}
    E_D=4\int_0^\infty\frac{dQ}{2\pi}\int_{-\infty}^0dv\,
    K(Q)\left[\mathcal W_Q^2+\mathcal W_v^2\right].
\end{equation}
The integration domain corresponds to the liquid region in one physical quadrant, while the factor of $4$ accounts for reflection in $x$ and $\tau$.

The physical action includes the empty region, where the action density is $U(0)$. When the liquid action is rewritten in hodograph variables on the instanton, a boundary term appears that cancels this vacuum contribution. This cancellation allows the total action to be expressed by $E_D$, whose integral covers only the liquid. Appendix~\ref{app:action-boundaries} gives the derivation and the regularity and endpoint conditions it requires.

The hodograph equation further reduces this two-dimensional integral to boundary values. Integrating by parts, we obtain
\begin{equation}\label{eq:dirichlet-energy}
    E_D = 4R\int_0^q\frac{dQ}{2\pi}\,K(Q)F(Q).
\end{equation}
Only the Neumann segment $v=0$, $0<Q<q$ contributes, where $\mathcal W=-F$ and $\mathcal W_v=-R$. The Dirichlet segment $v=0$, $Q>q$ gives no contribution because $\mathcal W=0$. The $Q=0$ boundary also does not contribute, since $K\mathcal W_Q=\pi\rho\tau\to0$ as $\rho\to0$ for finite imaginary time $\tau$.

Using $\mathcal I_{\mathrm{EFP}}=E_D$, Eq.~\eqref{eq:dirichlet-energy} gives the optimal action
\begin{equation}\label{eq:hydro-action-trace}
    \mathcal I_{\mathrm{EFP}}=\frac{2R}{\pi}\int_0^qK(Q)F(Q)\,dQ.
\end{equation}
Thus, the action requires only the values of $F(Q)=-\mathcal W(Q,0)$ on $[0,q]$.

For fixed finite Lieb--Liniger coupling, Appendix~\ref{app:ll-endpoint} shows that the remaining contour contribution near the slit endpoints $(\pm R,0)$ vanishes, assuming the stated regularity of the solution and thermodynamic coefficients. For fields that do not satisfy the hydrodynamic equations, the physical action is not simply $E_D$. The full transformed action, including its bulk and boundary terms, is derived in Appendix~\ref{app:action-offshell}.

In the next section, we construct the spectral modes $B_p(Q)$ by dressing with an auxiliary kernel and determine their amplitudes from an integral equation on a finite interval.

\section{The auxiliary-kernel method}\label{sec:auxiliary-kernel}
The spectral representation leaves two tasks: construct the modes $B_p(Q)$ satisfying Eq.~\eqref{eq:spectral-diffeq} and determine the amplitudes $a(p)$ in Eq.~\eqref{eq:spectral-superposition} from the mixed boundary conditions. We introduce the auxiliary-kernel method to address both. The idea is to represent the equation of state through an integral kernel $\Delta$, using formulas inspired by Bethe-ansatz thermodynamics. When such a representation is available, it generates the spectral modes and reduces the mixed boundary conditions to a single integral equation on a finite interval. Its solution determines the instanton density and current, the vacuum boundary, and the optimal action. The kernel is auxiliary: the model need not have an integrable microscopic description.

\subsection{Kernel thermodynamics and spectral modes}
We generalize the Lieb integral equation~\eqref{eq:problem-ll-thermodynamics} by replacing the Lieb--Liniger scattering kernel~\eqref{eq:ll-scattering-kernel} with a general real, even kernel $\Delta(\lambda-\mu)$. Consider a one-component Galilean Bethe-ansatz integrable model whose ground state fills a single rapidity interval $[-Q,Q]$ in the local rest frame~\cite{korepin1997quantum}. In this setting, $\Delta$ describes the two-body scattering shift.

The dressing of a bare function $f$ on $[-Q,Q]$ is defined by
\begin{equation}
    f_Q^\dr(\lambda) = f(\lambda) + \int_{-Q}^{Q} \frac{\Delta(\lambda-\mu)}{2\pi} f_Q^\dr(\mu) \, d\mu .
 \label{eq:dressing-equation}
\end{equation}
For a completely filled interval at zero temperature, the rapidity density $\rho_p(\lambda;Q)$ satisfies
\begin{equation}\label{eq:auxiliary-rapidity-density}
    \rho_p(\lambda;Q) = \frac{1}{2\pi}
    +\int_{-Q}^{Q}\frac{\Delta(\lambda-\mu)}{2\pi}\rho_p(\mu;Q)\,d\mu
    =\frac{1^\dr_Q(\lambda)}{2\pi}.
\end{equation}
The last equality follows by dressing the constant function $1$. Denote the finite-interval integral operator by
\begin{equation}\label{eq:finite-interval-operator}
    (\mathcal T_Q f)(\lambda)
    =\int_{-Q}^{Q}\frac{d\mu}{2\pi}\,
      \Delta(\lambda-\mu)f(\mu).
\end{equation}
We assume $1-\mathcal T_Q$ is invertible on the intervals of interest, so the dressing can be written as $f_Q^\dr=(1-\mathcal T_Q)^{-1}f$ on $[-Q,Q]$. We suppress the subscript $Q$ when the dressing interval is clear.

The rapidity density determines the moments
\begin{equation}\label{eq:charges}
    \mathcal Q[f]=\int_{-Q}^Q d\lambda\,\rho_p(\lambda;Q)f(\lambda), \qquad
    I_m=\mathcal Q[\lambda^m].
\end{equation}
In a Bethe integrable model, $I_m$ are the densities of conserved charges. In the local rest frame, all odd moments vanish. With bare dispersion $\varepsilon(\lambda)=\lambda^2/2$, the first two nonvanishing moments give the particle density $\rho=I_0$ and energy density $\epsilon=I_2/2$:
\begin{equation}
    \rho(Q) = \int_{-Q}^{Q}\rho_p(\lambda;Q) \, d\lambda, 
    \qquad
    \epsilon(Q) = \int_{-Q}^{Q} \frac{\lambda^2}{2}\rho_p(\lambda;Q) \, d\lambda.
\end{equation}
For a given kernel, both $\rho(Q)$ and $\epsilon(Q)$ depend on the interval endpoint $Q$. Since $\rho(Q)$ is increasing, as shown below, we can eliminate $Q$ to obtain the equation of state $\epsilon(\rho)$.

We will use these Bethe-ansatz formulas as an auxiliary representation of a prescribed equation of state. The kernel $\Delta$ must reproduce that equation of state, but need not have a microscopic scattering interpretation.

The derivation below assumes a regular kernel. Singular kernels require separate checks, which we carry out in Sec.~\ref{sec:polytropic} for polytropic gases with $1<\gpol\leq3$.

To connect this thermodynamic representation to the spectral problem, we differentiate the moments with respect to $Q$. These derivatives involve the endpoint value of the dressed constant function,
\begin{equation}\label{eq:dressed-charge}
    Z(Q)=1_Q^\dr(Q)=2\pi\rho_p(Q;Q).
\end{equation}
We assume $Z(Q)\ne0$, a continuously differentiable kernel, and the regularity of dressed derivatives and endpoint values specified in Appendix~\ref{app:dressed-modes}.

Because $\Delta(\lambda-\mu)=\Delta(\mu-\lambda)$, the operator $\mathcal T_Q$ and hence the dressing operator $(1-\mathcal T_Q)^{-1}$ are self-adjoint. Using $\rho_p=1_Q^\dr/(2\pi)$, we can therefore move the dressing from the constant function to $f$ in the moment integral:
\begin{align}\label{eq:charge-dressing-identity}
    \mathcal Q[f]
    &=\int_{-Q}^{Q}d\lambda\,\rho_p(\lambda;Q)f(\lambda) \nonumber \\
    &=\frac{1}{2\pi}\int_{-Q}^{Q}d\lambda\,1_Q^\dr(\lambda)f(\lambda) \nonumber \\
    &=\frac{1}{2\pi}\int_{-Q}^{Q}d\lambda\,f_Q^\dr(\lambda).
\end{align}
We now differentiate the last expression with respect to $Q$. The moving integration limits contribute $f_Q^\dr(Q)/\pi$, since the two endpoint values are equal. The change of the dressed function inside the interval contributes an additional $[Z(Q)-1]f_Q^\dr(Q)/\pi$. Adding these terms gives the endpoint identity for an even source $f(\lambda)$
\begin{equation}\label{eq:edge-identity}
    \frac{d}{dQ}\left[\frac{1}{2\pi}\int_{-Q}^Q f^\dr(\lambda)\, d\lambda\right] = \frac{Z(Q)f^\dr(Q)}{\pi}.
\end{equation}
The formula for an arbitrary source and its derivation are given in Eq.~\eqref{eq:app-general-moment-derivative} of Appendix~\ref{app:dressed-modes}.
Applying the moment and dressed endpoint identities to the density and energy then gives
\begin{equation}
    \frac{d\rho}{dQ}=\frac{Z^2}{\pi}, \qquad
    c_s=\frac{\pi\rho}{Z^2}, \qquad
    \frac{dQ}{d\rho}=\frac{c_s}{\rho}, \qquad
    K(Q)=Z(Q)^2.
\end{equation}
These relations identify $Q$ with the acoustic coordinate~\eqref{eq:acoustic-coord} and $Z^2$ with the local Luttinger parameter. For a regular kernel, $Z(0)=1$ and hence $K(0)=1$.\footnote{In the limit of vanishing density, $K$ approaches unity, the value for free spinless fermions. For the Lieb--Liniger gas, this is the well-known Tonks--Girardeau limit~\cite{korepin1997quantum}.}
Different vacuum scaling requires a singular kernel, as illustrated by polytropic gases in Sec.~\ref{sec:polytropic}.

To construct the spectral modes, consider how an even moment changes with density. Dividing Eq.~\eqref{eq:edge-identity} by $d\rho/dQ=Z^2/\pi$ gives
\begin{equation}
    \frac{d\mathcal Q[f]}{d\rho} = \frac{f^\dr(Q)}{Z(Q)}.
\end{equation}
Choose the even source $f_p(\lambda)=\cos(p\lambda)$. Its moment $\mathcal Q_p=\mathcal Q[f_p]$ generates the even rapidity moments. Since $f_p''=-p^2f_p$, we use its density derivative to construct a mode with spectral parameter $p$:
\begin{equation}\label{eq:dressed-spectral-mode}
    B_p(Q) = \frac{d\mathcal Q_p}{d\rho} = \frac{[\cos(p\lambda)]^\dr(Q)}{Z(Q)}.
\end{equation}
To differentiate $B_p$, we use the endpoint identities~\eqref{eq:app-even-edge-derivative} and~\eqref{eq:app-odd-edge-derivative} for even and odd sources, derived in Appendix~\ref{app:dressed-modes}. The first derivative of $B_p$ brings in the dressed sine function, while the second returns the cosine and produces the term proportional to $K'/K$, with $K=Z^2$. The result is
\begin{equation}
    \frac{1}{K}\partial_Q(K\partial_QB_p)=-p^2B_p.
\end{equation}
Thus, from an auxiliary kernel representing the equation of state, we have constructed the modes $B_p(Q)$ satisfying the hydrodynamic spectral equation~\eqref{eq:spectral-diffeq}. For a regular kernel, the modes also satisfy $B_p(0)=1$ and $B_p'(0)=0$ at the vacuum boundary $Q = 0$, as shown in Appendix~\ref{app:dressed-modes}. It remains to determine the amplitudes $a(p)$ from the mixed boundary conditions.

\subsection{Resolvent formulation of the mixed boundary problem}\label{sec:compact-reconstruction}
Since each mode satisfies $B_p'(0)=0$, the expansion~\eqref{eq:spectral-superposition} already satisfies the vacuum condition $\mathcal W_Q(0,v)=0$ for $v<0$. We seek a solution that tends to zero as $Q^2+v^2\to\infty$. To determine the amplitudes $a(p)$, we impose the Dirichlet and Neumann conditions~\eqref{eq:hodograph-dirichlet} and~\eqref{eq:hodograph-neumann} on $v=0$, repeated here for convenience:
\begin{equation}\label{eq:mixed-boundary-data}
    \mathcal W(Q,0)=0\quad(Q>q),\qquad
    \mathcal W_v(Q,0)=-R\quad(0<Q<q).
\end{equation}
Here, $q$ is the fixed background value of the acoustic coordinate, whereas $Q$ labels the local fluid state. We will parametrize the amplitudes $a(p)$ by an unknown function $\chi(k)$ on $[-q,q]$, held fixed as $Q$ varies. This will enforce the Dirichlet condition automatically; the Neumann condition will then determine $\chi$.

\paragraph*{Dirichlet condition, $Q>q$.} 
For a function $\varphi(k)$, define the whole-line convolution operator using the same kernel:
\begin{equation}\label{eq:convolution-def}
    (\mathcal T_\infty\varphi)(k)
    =\int_{\mathbb R}\frac{dt}{2\pi}\,
    \Delta(t)\varphi(k-t),\qquad
    \widehat{\mathcal T_\infty\varphi}(p)
    =\kappa(p)\widehat\varphi(p),
\end{equation}
with the Fourier convention
$\widehat\varphi(p)=\int_{\mathbb R}e^{-ipk}\varphi(k)\,dk$.
Here $\kappa(p)$ is the normalized Fourier transform of $\Delta$:
\[
    \kappa(p)=\int_{\mathbb R}\frac{dt}{2\pi}\,e^{-ipt}\Delta(t)
    =\frac{\widehat\Delta(p)}{2\pi}.
\]
Since $\Delta$ is real and even, so is $\kappa$. We choose $\chi$ real and even, consistent with the reflection symmetry of the kernel and boundary data, and set
\begin{equation}\label{eq:chi-spectral-amplitude}
    a(p)=\frac{1-\kappa(p)}{\pi}\widehat\chi(p),\qquad
    \widehat\chi(p)=\int_{-q}^q\chi(k)\cos(pk)\,dk.
\end{equation}
We assume that $\chi$ is sufficiently regular, vanishes at $k=\pm q$, and is extended by zero outside $[-q,q]$. The precise regularity and operator assumptions used below are stated in Appendix~\ref{app:resolvent-equivalence}. The factor $1-\kappa$ is chosen to cancel the dressing on the Dirichlet segment, as we now verify.

Substituting these amplitudes into the spectral representation~\eqref{eq:spectral-superposition} and using the linearity of dressing and the identity $\mathcal T_\infty\cos(pk)=\kappa(p)\cos(pk)$, we obtain, for $v<0$,
\begin{equation}\label{eq:chi-spectral-reconstruction}
    \mathcal W(Q,v)=-\frac{1}{Z(Q)}
    \left[(1-\mathcal T_\infty)\int_0^\infty\frac{dp}{\pi}\,
    \widehat\chi(p)e^{pv}\cos(pk)\right]^\dr_Q(Q).
\end{equation}
Whole-line convolution is performed first, followed by dressing on $[-Q,Q]$ and evaluation at $k=Q$. 

As $v\to0^{-}$, the Fourier integral in Eq.~\eqref{eq:chi-spectral-reconstruction} recovers $\chi(k)$ by Fourier inversion. For $Q\geq q$, the interval $[-Q,Q]$ contains the whole support of $\chi$. Therefore, for $|k|\leq Q$,
\begin{equation*}
    (\mathcal T_\infty\chi)(k)
    =\int_{-q}^{q}\frac{d\mu}{2\pi}\,\Delta(k-\mu)\chi(\mu)
    =(\mathcal T_Q\chi)(k).
\end{equation*}
Consequently, the function to be dressed at $v=0$ is $(1-\mathcal T_Q)\chi$ on $[-Q,Q]$. Dressing applies $(1-\mathcal T_Q)^{-1}$ and returns $\chi$. Evaluating at $k=Q$ gives
\begin{equation*}
    \mathcal W(Q,0) = -\frac{[(1-\mathcal T_Q)\chi]^\dr_Q(Q)}{Z(Q)} = -\frac{\chi(Q)}{Z(Q)}=0,\qquad Q>q.
\end{equation*}
Here $\chi(Q)=0$ because $Q>q$, so the Dirichlet condition is satisfied.

\paragraph*{Cauchy transform.} To impose the Neumann condition, we express the Fourier integral in Eq.~\eqref{eq:chi-spectral-reconstruction} as the imaginary part of an analytic function. Its real part will then determine the $v$ derivative through the Cauchy--Riemann relations. Define the Cauchy transform, or resolvent, of $\chi$ by
\begin{equation}\label{eq:resolvent-definition}
    C(z)=\frac{1}{\pi R}\int_{-q}^q\frac{\chi(k)}{z-k}\,dk.
\end{equation}
The resolvent is analytic off $[-q,q]$. Reality and evenness of $\chi$ give $C(\bar z)=\overline{C(z)}$ and $C(-z)=-C(z)$. Its boundary values $C_\pm(k)=C(k\pm i0)$ recover $\chi$ through $\operatorname{Im}C_+=-\chi/R$ or, equivalently, $C_+-C_-=-2i\chi/R$. In the upper half-plane, write $z=k-iv$ with $v<0$. Then
\begin{equation*}
    \operatorname{Im}C(k-iv)=-\frac{1}{\pi R}
    \int_0^\infty\widehat\chi(p)e^{pv}\cos(pk)\,dp.
\end{equation*}
This reproduces the Fourier integral in Eq.~\eqref{eq:chi-spectral-reconstruction}, with the factor $-1/R$. To include the auxiliary kernel, define
\begin{equation}\label{eq:kernel-resolvent}
    C_\Delta(z)=(1-\mathcal T_\infty)C(z)
    =C(z)-\int_{\mathbb R}\frac{dt}{2\pi}\,\Delta(t)C(z-t),
\end{equation}
The reconstruction~\eqref{eq:chi-spectral-reconstruction} now takes the form
\begin{equation}\label{eq:resolvent-reconstruction}
    \mathcal W(Q,v)
    =R\,\frac{[\operatorname{Im}C_\Delta(k-iv)]^\dr_Q(Q)}{Z(Q)}.
\end{equation}
The imaginary part $\operatorname{Im}C_+(k)=-\chi(k)/R$ vanishes outside $[-q,q]$. The real part generally does not. The convolution $\mathcal T_\infty C_+$ must therefore include the values of $\operatorname{Re}C_+(k)$ outside this interval.

\paragraph*{Neumann condition, $0<Q<q$.} At fixed $Q$, the velocity derivative of the reconstructed potential acts only on $C_\Delta$. The Cauchy--Riemann relation $\partial_v\operatorname{Im}C_\Delta(k-iv)=-\partial_k\operatorname{Re}C_\Delta(k-iv)$ gives
\begin{equation}\label{eq:resolvent-neumann}
    \mathcal W_v(Q,0^-)
    =-R\,\frac{[\operatorname{Re}C'_{\Delta,+}]^\dr_Q(Q)}{Z(Q)}.
\end{equation}
To obtain $\mathcal W_v=-R$ for every $0<Q<q$, it is enough to require   $\operatorname{Re}C'_{\Delta,+}(k)=1$ for $|k|<q$. Since $[-Q,Q]\subset(-q,q)$,   dressing $\operatorname{Re}C'_{\Delta,+}$ on $[-Q,Q]$ is equivalent to dressing the  constant function $1$. Its dressed endpoint value is $1_Q^\dr(Q)=Z(Q)$, so 
\begin{equation*}
    \mathcal W_v(Q,0^-)
    =-R\,\frac{1_Q^\dr(Q)}{Z(Q)}
    =-R\,\frac{Z(Q)}{Z(Q)}
    =-R,\qquad 0<Q<q.
\end{equation*}
Integrating $\operatorname{Re}C'_{\Delta,+}=1$ with respect to $k$ gives $\operatorname{Re}C_{\Delta,+}(k)=k+\mathrm{const}$. Oddness of $\operatorname{Re}C_{\Delta,+}$ fixes the constant to zero. Using Eq.~\eqref{eq:kernel-resolvent}, we obtain
\begin{equation}\label{eq:resolvent-boundary}
    \operatorname{Re}C_{\Delta,+}(k)
    =(1-\mathcal T_\infty)\operatorname{Re}C_+(k)=k,
    \qquad |k|<q.
\end{equation}
The Neumann condition for $\mathcal W$ has therefore become a Dirichlet condition for $\operatorname{Re}C_\Delta$ on $(-q,q)$. Together with the Cauchy representation~\eqref{eq:resolvent-definition}, this is an integral equation for $\chi$. Under the stated regularity assumptions, Appendix~\ref{app:resolvent-equivalence} shows that this resolvent condition is also necessary for the Neumann boundary condition.

In electrostatic terms, $\chi$ acts as a line-charge density and $C$ as its complex field. The auxiliary kernel modifies this field, which balances an applied linear field corresponding to a quadratic external potential (see Appendix~\ref{app:integrated-equation}).

\begin{figure}[htbp]
    \centering
    \includegraphics[width=\linewidth]{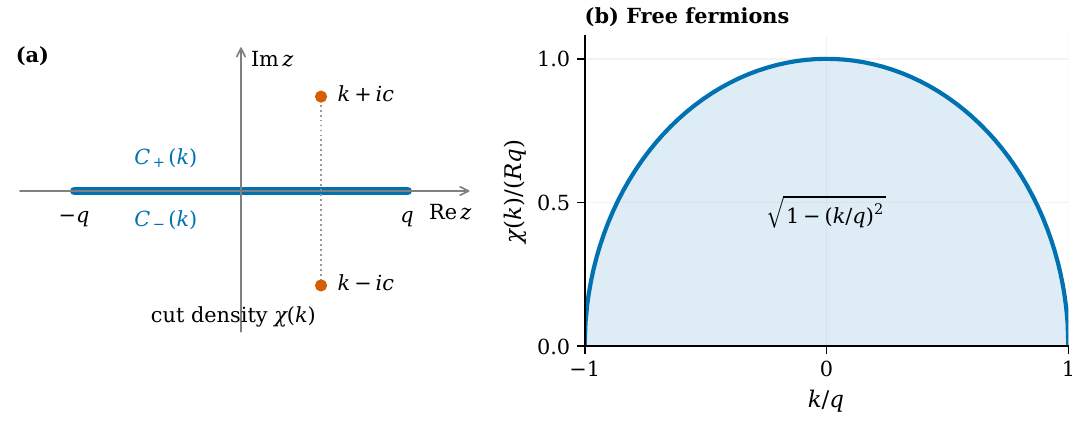}
    \caption{The Cauchy representation. (a) The unknown $\chi$ is supported on the cut $[-q,q]$. Its jump fixes $C_+-C_-=-2i\chi/R$. The points $k\pm ic$ indicate the shifted values appearing for the Lieb--Liniger kernel in Eq.~\eqref{eq:hoppe-ll-shift}. (b) The exact free-fermion auxiliary density $\chi(k)/(Rq)=\sqrt{1-(k/q)^2}$.}
    \label{fig:cauchy-resolvent}
\end{figure}

For free fermions, $\Delta=0$ and the equation has the elementary solution (cf.~Ref.~\cite{abanov-hydro})
\begin{equation}\label{eq:free-fermion-resolvent}
    C(z)=z-\sqrt{z^2-q^2},\qquad
    \chi(k)=R\sqrt{q^2-k^2}\quad (|k|<q),
\end{equation}
where the square root is asymptotic to $z$ at infinity. On the cut, $\operatorname{Re}C_+(k)=k$, and $C(z)\sim q^2/(2z)$. Figure~\ref{fig:cauchy-resolvent} illustrates this representation. The coefficient of $1/z$ will be used to determine the action below.

\paragraph*{A real integral equation.} The real part of the Cauchy transform involves the principal-value kernel $1/(k-\mu)$. Integrating by parts replaces this kernel by a logarithm and moves the derivative onto $\chi$. This motivates introducing the normalized derivative
\begin{equation}\label{eq:mathfrak-g-def}
    \mathfrak g(k)=-\frac{2\chi'(k)}{\pi R}
    =\frac2\pi\frac{d}{dk}\operatorname{Im}C_+(k).
\end{equation}
The function $\mathfrak g$ is odd and integrable. Because $\chi(\pm q)=0$, the boundary term in the integration by parts vanishes, so
\begin{equation*}
    \begin{aligned}
    \operatorname{Re}C_+(k)
    &=\frac{1}{\pi R}\operatorname{PV}\!\int_{-q}^q
      \frac{\chi(\mu)}{k-\mu}\,d\mu\\
    &=-\frac12\int_{-q}^q\mathfrak g(\mu)
      \log\frac{|k-\mu|}{q}\,d\mu.
    \end{aligned}
\end{equation*}
Applying $1-\mathcal T_\infty$ to this expression modifies the logarithmic kernel by convolution with $\Delta$. Where this convolution converges, write $L(x)=-2(1-\mathcal T_\infty)\log(|x|/q)$. The resolvent boundary condition~\eqref{eq:resolvent-boundary} then gives
\begin{equation}\label{eq:mathfrakg}
    \int_{-q}^q L(k-\mu)\mathfrak g(\mu)\,d\mu=4k,
    \qquad |k|<q.
\end{equation}
More generally, the even kernel $L$ is defined by its Fourier transform,
\begin{equation}\label{eq:integrated-kernel}
    \widehat L(p)=\frac{2\pi[1-\kappa(p)]}{|p|}.
\end{equation}
An additive constant in the logarithm or in $L$ has no effect because $\int_{-q}^q\mathfrak g(\mu)\,d\mu=0$. The detailed derivation is given in Appendix~\ref{app:integrated-equation}.

Once $\mathfrak g$ is found, integration recovers $\chi$, with the constant fixed by $\chi(q)=0$, while the oddness of $\mathfrak g$ ensures $\chi(-q)=0$ as well. The Cauchy transform then gives $C$, and Eq.~\eqref{eq:resolvent-reconstruction} reconstructs $\mathcal W$. The normalization cancels $R$ from the equations for $C$ and $\mathfrak g$, which depend only on the kernel and $q$. The resolvent and real integral equations are equivalent under the endpoint and operator assumptions stated above. Existence and uniqueness require additional conditions, discussed in Appendix~\ref{app:compact-uniqueness}.

For singular kernels, the whole-line operators need a Fourier or distributional interpretation, and the reconstruction, vacuum limit, and action must be checked separately. This is done explicitly for the polytropic family in Section~\ref{sec:polytropic}.

\subsection{Vacuum boundary and action from the resolvent}\label{sec:response-observables}
For a given auxiliary kernel $\Delta$, the resolvent $C$ together with the spectral modes $B_p(Q)$ determines the full hydrodynamic instanton. Remarkably, the vacuum boundary and optimal action follow from $C$ alone, without reconstructing the full liquid profile.

\paragraph*{Vacuum boundary.} We continue to assume the regular-kernel conditions of Sec.~\ref{sec:compact-reconstruction} and the vacuum-boundary conditions of Sec.~\ref{sec:hodograph-problem}. In the reconstruction~\eqref{eq:resolvent-reconstruction}, taking $Q\to0$ at fixed $v<0$ makes finite-interval dressing tend to the identity and $Z(Q)\to1$. The potential on the vacuum boundary and the characteristic intercept $X(v)=R+\partial_v\mathcal W(0,v)$ are therefore
\begin{equation*}
    \mathcal W(0,v)=R\operatorname{Im}C_\Delta(-iv),\qquad
    X(v)=R[1-\operatorname{Re}C'_\Delta(-iv)].
\end{equation*}
To express the curve in dimensionless form, introduce $\eta=|v|/q>0$ and write $X(-q\eta)=Rb_\Delta(\eta)$, where
\begin{equation}\label{eq:general-boundary-b}
    b_\Delta(\eta)=1-\operatorname{Re}C'_\Delta(iq\eta).
\end{equation}
Since $X'(v)=-(R/q)b'_\Delta(\eta)$, the envelope relations~\eqref{eq:vacuum-envelope} give the upper-right vacuum boundary explicitly:
\begin{equation}\label{eq:general-vacuum-boundary}
    \frac{x(\eta)}{R}=b_\Delta(\eta)-\eta b'_\Delta(\eta),
    \qquad
    \frac{q\tau(\eta)}{R}=b'_\Delta(\eta),
    \qquad \eta>0.
\end{equation}
Here $C'_\Delta$ denotes differentiation in $z$, while $b'_\Delta$ denotes differentiation in $\eta$. The endpoint values are understood as limits from the liquid side, and reflections in $x$ and $\tau$ complete the curve.

If we solve the real integral equation~\eqref{eq:mathfrakg}, we can compute $X(v)$ directly from $\mathfrak g$. Set $Q=0$ in the spectral representation~\eqref{eq:spectral-superposition}, where $B_p(0)=1$, and substitute the amplitudes~\eqref{eq:chi-spectral-amplitude}. Integration by parts in the Fourier transform of $\chi$ replaces it by its derivative, which defines $\mathfrak g$, and the boundary term vanishes since $\chi(\pm q)=0$. Differentiating with respect to $v$ then gives the intercept $X(v)=R+\partial_v\mathcal W(0,v)$ in the form
\begin{equation}\label{eq:main-general-X}
    X(v)=R\left[1-\int_0^q dk\,\mathfrak g(k)
    \int_0^\infty dp\,[1-\kappa(p)]e^{pv}\sin(pk)\right],
    \qquad v<0.
\end{equation}
\paragraph*{Action.} The boundary integral for the action~\eqref{eq:hydro-action-trace} depends on $F(Q)=-\mathcal W(Q,0)$. We can eliminate this function using the identity
\begin{equation*}
    2\int_0^q K(Q)F(Q)\,dQ=\int_{-q}^q\chi(k)\,dk.
\end{equation*}
This identity follows by varying the dressing endpoint $Q$ at fixed $\Delta$, $q$, and $\chi$ and integrating from $Q=0$ to $Q=q$. The derivation and its assumptions are given in Appendix~\ref{app:action-boundary-values}.

Using this identity in Eq.~\eqref{eq:hydro-action-trace}, we obtain $\mathcal I_{\rm EFP}=(R/\pi)\int_{-q}^q\chi(k)\,dk$. Expanding the Cauchy transform~\eqref{eq:resolvent-definition} at large $z$ shows that the same integral determines its leading coefficient. Thus, 
\begin{equation}\label{eq:resolvent-action}
    C(z)=\frac A z+O(z^{-3}),\qquad
    \mathcal I_{\rm EFP}=R^2A,\qquad
    A=\frac1{\pi R}\int_{-q}^q\chi(k)\,dk.
\end{equation}
Cf.~Ref.~\cite[Sec.~5]{abanov-hydro} for an analogous relation between the action and the instanton's far-field asymptotics.
Using $\chi(\pm q)=0$, integration by parts expresses the same action through the first moment of $\mathfrak g$:
\begin{equation}\label{eq:general-response-action}
    \mathcal I_{\mathrm{EFP}}
    =\frac{R^2}{2}\int_{-q}^{q} k\,\mathfrak g(k)\,dk.
\end{equation}
In the $(qR)^2$ normalization of Eq.~\eqref{eq:rescaled-rate-definition}, the rate is
\begin{equation}\label{eq:main-general-response}
    \widehat f=\frac{\mathcal I_{\mathrm{EFP}}}{(qR)^2}
    =\frac{1}{2q^2}\int_{-q}^{q}k\mathfrak g(k)\,dk.
\end{equation}
For the Lieb--Liniger gas, substituting $k=qt$ and $g(t)=\mathfrak g(qt)$ recovers Eq.~\eqref{eq:main-ll-rate}.

\subsection{Applying the auxiliary-kernel method}\label{sec:practical-procedure}
The auxiliary-kernel construction gives the following steps for solving the EFP problem in a Galilean fluid.

\begin{list}{\textbf{\arabic{enumi}.}}{%
    \usecounter{enumi}%
    \setlength{\leftmargin}{1.5em}%
    \setlength{\labelwidth}{1em}%
    \setlength{\labelsep}{0.5em}%
    \setlength{\rightmargin}{0pt}%
    \setlength{\itemsep}{0.6\baselineskip}%
    \setlength{\topsep}{0.5\baselineskip}%
    \setlength{\parsep}{0pt}%
    \setlength{\partopsep}{0pt}%
}
\item We start with the energy density $\epsilon(\rho)$ and the background density $\rho_0$. These determine other thermodynamic quantities, such as the sound speed  $c_s(\rho)$, the acoustic coordinate $Q(\rho)$, and the Luttinger parameter (weight) $K(Q)$, as described in Sec.~\ref{sec:hodograph-problem}. 

\item Next, we find a real, even kernel $\Delta$ that reproduces the equation of state over the required density range. Such a kernel need not come from an integrable microscopic model. When it exists, dressing gives the spectral modes $B_p(Q)$ through  Eq.~\eqref{eq:dressed-spectral-mode}. Their decaying superposition~\eqref{eq:spectral-superposition} satisfies $\mathcal W_Q(0,v)=0$, leaving the other boundary conditions to determine the amplitudes $a(p)$.

\item For an empty interval $[-R,R]$, we solve Eq.~\eqref{eq:mathfrakg} for an odd, integrable function $\mathfrak g$ on $[-q,q]$. This can be done analytically or numerically. We then recover $\chi$ using Eq.~\eqref{eq:chi-from-g}, with $\chi(\pm q)=0$.

\item Once $\mathfrak g$ is known, the action follows from the  moment~\eqref{eq:general-response-action}. The same function determines $X(v)$ through Eq.~\eqref{eq:main-general-X}, and hence the vacuum boundary as the envelope of the lines $x-v\tau=X(v)$, using Eq.~\eqref{eq:vacuum-envelope}. Neither calculation requires the spectral modes or the full liquid profile.

\item To find the density and current throughout the instanton, we first construct $C$ and $C_\Delta$ from $\chi$ using Eqs.~\eqref{eq:resolvent-definition} and~\eqref{eq:kernel-resolvent}. Dressing on the finite interval then gives $\mathcal W(Q,v)$ through Eq.~\eqref{eq:resolvent-reconstruction}. Finally, the inverse map~\eqref{eq:hodograph-mapping} gives the spacetime coordinates $(x,\tau)$ corresponding to the density $\rho(Q)$ and current $\rho(Q)v$.
\end{list}

The kernel representation and construction of spectral modes are not specific to the EFP, and can also be used for hydrodynamic problems with different boundary conditions. With the same condition at $Q=0$, the modes $B_p(Q)$ also give separated solutions of the hodograph equation for smooth real-time dynamics. In real time, the equation is hyperbolic, and the velocity dependence is oscillatory rather than decaying. Other applications require appropriate initial or boundary data and a separate check that the hodograph map remains invertible. The integral equation for $\mathfrak g$ and the action formulas above, however, are specific to the EFP boundary conditions.

We now apply the EFP procedure to the repulsive Lieb--Liniger gas.

\section{Lieb--Liniger gas and numerical comparison}\label{sec:ll-application}
For the repulsive Lieb--Liniger gas, the physical scattering kernel~\eqref{eq:ll-scattering-kernel} provides the auxiliary representation required. We first obtain the integral equation and explain how its solution gives the rate, vacuum boundary, and full instanton. For this particular kernel, the equation admits an explicit elliptic solution: its rescaled and normalized resolvent form coincides with the equation solved by KKN~\cite{kazakov1999dparticles}. We present this solution and compare the hydrodynamic predictions with direct numerical minimization of the action.

\subsection{The Lieb--Liniger integral equation}\label{sec:ll-integral-equation}
For a given density $\rho_0$ and physical coupling $\gamma_0=c/\rho_0$, the Lieb equation~\eqref{eq:problem-ll-thermodynamics} determines the background Fermi rapidity $q$. The scattering kernel $\Delta(\lambda)$~\eqref{eq:ll-scattering-kernel}, its normalized Fourier transform $\kappa=\hat\Delta/2\pi$, and the resulting kernel $L$ in Eq.~\eqref{eq:mathfrakg} are
\begin{equation}\label{eq:ll-kernel-multipliers}
    \Delta(\lambda)=\frac{2c}{c^2+\lambda^2},\qquad
    \kappa(p)=e^{-c|p|},\qquad L(\lambda)=\log\left(1+\frac{c^2}{\lambda^2}\right).
\end{equation}
Rescaling rapidities by $q$, with $r=c/q$ and $g(t)=\mathfrak g(qt)$, reduces the general equation~\eqref{eq:mathfrakg} to Eq.~\eqref{eq:main-ll-response}. We seek an odd solution with integrable inverse-square-root behavior at $t=\pm1$. Equivalently, $\chi$ is even and vanishes as a square root at $k=\pm q$.

The Lieb--Liniger kernel also simplifies the resolvent reconstruction. Whole-line convolution shifts the argument of the Cauchy transform: $\mathcal T_\infty C(z)=C(z+ic)$ for $\operatorname{Im}z>0$. Hence $C_\Delta(z)=C(z)-C(z+ic)$. Both $g$ and $C$ are fixed by the background parameters $c$ and $q$.

\subsection{Rate, vacuum boundary, and density profile}\label{sec:ll-reconstruction}
Once $g$ is known, the formulas of Sec.~\ref{sec:response-observables} determine the observables. The reconstruction below applies whether $g$ is obtained numerically or analytically.

\paragraph{Rate.}
Rescaling the action moment~\eqref{eq:general-response-action} by $q$ gives $\mathcal I_{\mathrm{EFP}}=(qR)^2\widehat f(r)$ and the first-moment formula~\eqref{eq:main-ll-rate}. This derives the conjectured rate of Ref.~\cite{khanikati2026efp}, under the physical and boundary assumptions stated in Secs.~\ref{sec:macroscopic-limit} and~\ref{sec:hydro-observables}.

\paragraph{Vacuum boundary.}
The exponential form of $\kappa$ in Eq.~\eqref{eq:ll-kernel-multipliers} makes the inner integral for the characteristic intercept~\eqref{eq:main-general-X} elementary. With $y=|v|>0$,
\begin{equation*}
    \int_0^\infty dp\,(1-e^{-cp})e^{-py}\sin(p\lambda)
    =\frac{\lambda}{\lambda^2+y^2}
     -\frac{\lambda}{\lambda^2+(y+c)^2}.
\end{equation*}
Putting $\lambda=qt$ and $y=q\eta$ gives $b_\Delta(\eta)=b(\eta)$, with $b$ defined in Eq.~\eqref{eq:main-ll-b}. This function and its derivative give the upper-right arc through Eq.~\eqref{eq:main-ll-boundary}, and reflections complete the boundary. 

\paragraph{Liquid profile.}
Starting from $g$, we recover $\chi$ by integration, using $\mathfrak g(k)=g(k/q)$ in Eq.~\eqref{eq:mathfrak-g-def} and the endpoint condition in Eq.~\eqref{eq:chi-from-g}. We then form its Cauchy transform $C$. The potential $\mathcal W$ follows by dressing the imaginary part of $C_\Delta$ according to Eq.~\eqref{eq:resolvent-reconstruction}.

This dressing can be evaluated using the Lieb--Liniger ground-state rapidity distributions $\rho_p(k;Q)=1_Q^\dr(k)/(2\pi)$, normalized by $\rho(Q)=\int_{-Q}^Q\rho_p(k;Q)\,dk$. 
Once this thermodynamic family is known, evaluating $\mathcal W$ at different velocities requires only quadrature and differentiation, without solving a new dressing equation for each velocity. For $y=-v>0$, we obtain
\begin{equation}\label{eq:ll-profile-quadrature}
    \mathcal W(Q,-y)=\frac{R}{\rho'(Q)}\,\partial_Q
    \int_{-Q}^Q\rho_p(k;Q)\,
    \operatorname{Im}\!\left[C(k+iy)-C(k+i(y+c))\right]dk.
\end{equation}
Here $\rho'(Q)=Z(Q)^2/\pi$. The derivative acts on both the integration limits and the distribution $\rho_p(k;Q)$, while $q,c,y$, and the background resolvent $C$ remain fixed. Appendix~\ref{app:hoppe-profile-quadrature} derives this identity.

The hodograph map~\eqref{eq:hodograph-mapping} then gives $(x,\tau)$ for each $(Q,v)$, with density $\rho(Q)$ and current $\rho(Q)v$. Reflections in $x$ and $\tau$ complete the instanton configuration.

\subsection{Applying the elliptic solution of KKN}\label{sec:ll-elliptic-rate}
KKN's elliptic solution gives both $g$ and the shifted resolvent $C_\Delta$ explicitly. The connection follows from the shift property in Sec.~\ref{sec:ll-integral-equation}. Together with Schwarz symmetry, it puts the general resolvent boundary condition~\eqref{eq:resolvent-boundary} in the form
\begin{equation}\label{eq:hoppe-ll-shift}
    C_+(k)+C_-(k)-C(k+ic)-C(k-ic)=2k,
    \qquad |k|<q.
\end{equation}
After rescaling rapidity by $c$ and normalizing the resolvent at infinity, this is the equation solved by KKN~\cite[Sec.~6.1]{kazakov1999dparticles} in their treatment of the Hoppe model. The normalization and the dictionary between the two problems are presented in Appendix~\ref{app:hoppe-normalization}.

For each coupling $r>0$, the solution is parametrized by a unique real elliptic parameter $m\in(0,1)$, related to $r$ by Eq.~\eqref{eq:ll-elliptic-r}. In our variables, it gives explicit expressions for the rate $\widehat f(r)$ in Eq.~\eqref{eq:ll-elliptic-rate} and for the auxiliary functions $\chi$ and $g$ in Eqs.~\eqref{eq:hoppe-elliptic-chi}--\eqref{eq:hoppe-elliptic-g}.

The same solution gives the entire vacuum boundary through Eq.~\eqref{eq:hoppe-elliptic-vacuum}, without integrating $g$. In particular, the vacuum closes on the symmetry axis at
\begin{equation}\label{eq:ll-elliptic-closing}
    \tau_c=\frac{2R}{c}\left[\frac{m}{(1-\mathbf E(m)/\mathbf K(m))^2}-1\right],
\end{equation}
where $\mathbf K$ and $\mathbf E$ are the complete elliptic integrals of the first and second kinds, respectively. The solution recovers both limiting vacuum curves, with $q\tau_c/R\to1$ for free fermions and $q\tau_c/R\to4/\pi$ in the weak-Bose limit. The corresponding rates are $\widehat f\to1/2$ and $\widehat f\sim2/(3\pi r)$, as shown in Fig.~\ref{fig:ll-efp-rate}.

For the liquid profile, Eq.~\eqref{eq:hoppe-elliptic-Cdelta} gives $C_\Delta$ directly, so there is no need to reconstruct $\chi$ and $C$ from $g$. The quadrature~\eqref{eq:ll-profile-quadrature} still requires the Lieb--Liniger distributions $\rho_p(k;Q)$. Together with $C_\Delta$, these determine the full instanton through the reconstruction of Sec.~\ref{sec:ll-reconstruction}. 
Appendix~\ref{app:hoppe-elliptic} collects the formulas in our normalization and explains how to evaluate them, including the branch choices, endpoint conditions, and coupling limits.

\begin{figure}[htbp]
    \centering
    \includegraphics[width=0.88\linewidth]{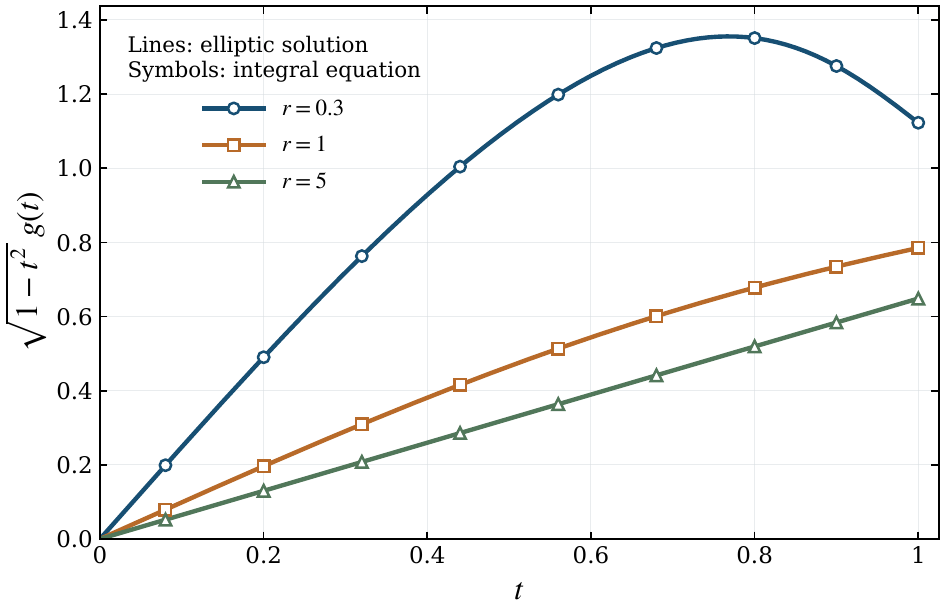}
    \caption{The function $\sqrt{1-t^2}\,g(t)$ at three couplings $r=c/q$. Solid curves show the elliptic solution \eqref{eq:hoppe-elliptic-g}, and symbols show numerical solutions of Eq.~\eqref{eq:main-ll-response}. The factor $\sqrt{1-t^2}$ removes the endpoint divergence of $g$, and values at $t=1$ are understood as limits. Only the positive half of the interval is shown, since $g$ is odd.}
    \label{fig:ll-hoppe-g}
\end{figure}

Figure~\ref{fig:ll-hoppe-g} illustrates the coupling dependence of $g$ and compares the elliptic solution with a direct numerical solution of the same integral equation.

\subsection{Comparison with numerical minimization}\label{sec:hydro-numerics}
We test the predicted vacuum geometry by numerically minimizing the Euclidean action. A displacement field $\mathcal U$ enforces continuity through $\rho=\mathcal U_x$ and the corresponding current, while the constraint $\rho\geq0$ allows a vacuum region to emerge.
Details of the discretization, high-density continuation, and convergence checks are presented in Appendix~\ref{app:hydro-numerics}.

Figure~\ref{fig:instanton-density-profiles} shows the spacetime density profiles obtained from these numerical minimizations. The density is strongly depleted inside the predicted arctic curve and diverges near the slit endpoints. The finite optimization barrier leaves a small positive density in the nominal vacuum. The gold curves show the analytic vacuum boundaries. A quantitative comparison with the numerically extracted interfaces is given in Fig.~\ref{fig:vacuum-boundary-comparison} and Table~\ref{tab:hydro-convergence}.


\begin{figure}[t]
    \centering
    \includegraphics[width=\linewidth]{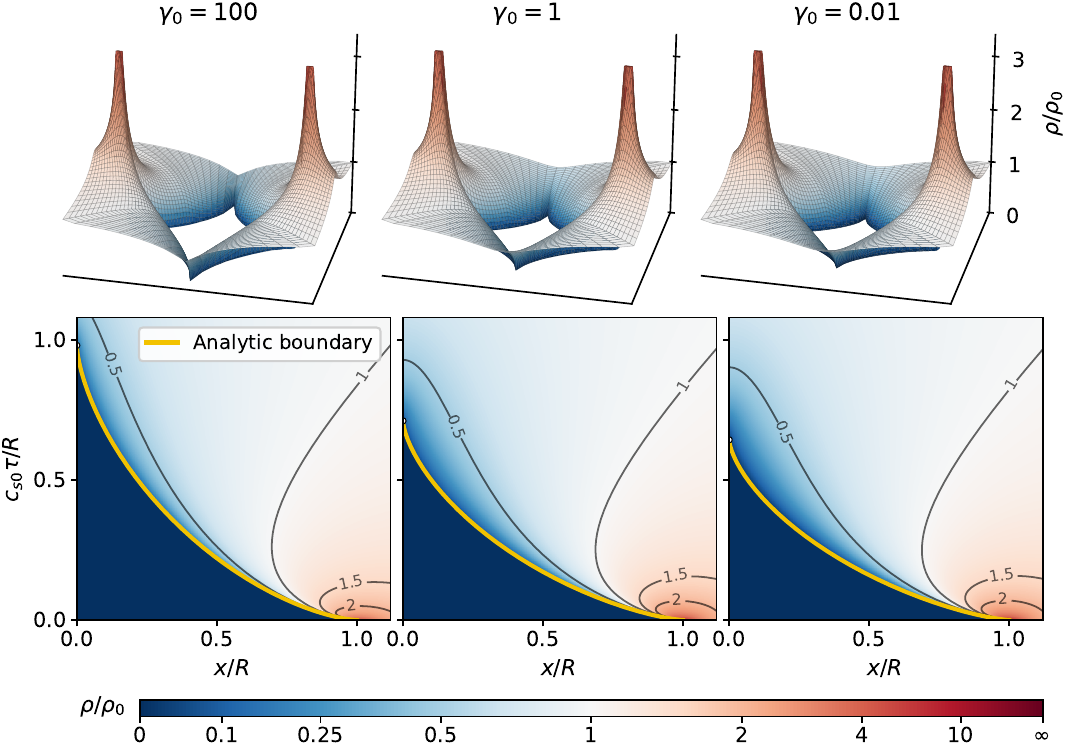}
    \caption{Numerically minimized density $\rho/\rho_0$ for $\gamma_0=100,1,0.01$. Top: full spacetime view of instantons, with density clipped at approximately $3\rho_0$ near the divergent slit endpoints. Bottom: upper-right quadrant, colored by $\rho/(\rho+\rho_0)$, with contours at $\rho/\rho_0=0.5,1,1.5,2$. Gold curves are the analytic vacuum boundaries, and open circles mark the numerical closing times.}
    \label{fig:instanton-density-profiles}
\end{figure}

We extract the vacuum boundary by extrapolating the chemical potential from the liquid side, obtaining the boundary points in Fig.~\ref{fig:vacuum-boundary-comparison}. Neither the vacuum boundary nor its closing time is imposed. The outer boundary  condition fixes the far-field dipole amplitude to its analytic value, so the comparison tests the interior geometry rather than the amplitude.

For $\gamma_0=\infty,100,10,1,0.01$, the fitted closing time changes by less than $0.01\%$ across the last three refinements. The root-mean-square (RMS) difference between the numerical and analytic boundaries is of order $10^{-3}R$, with larger deviations near the closing point. The cusp regions are not fully resolved and are excluded from the comparison. Overall, the numerical results agree well with the predicted vacuum geometry. On the finest meshes, the corrected numerical actions are $0.007$--$0.023$\% above the analytic prediction. These values decrease with each refinement, although convergence to a mesh-independent value has not yet been established.

\section{Polytropic gas}\label{sec:polytropic}
For polytropic fluids, we choose an auxiliary kernel to reproduce a prescribed equation of state, without assuming an integrable microscopic model. Their EFP instantons are known from a direct hydrodynamic solution~\cite{abanov2025polytropicEFP}, which allows us to check the construction. We first summarize the hydrodynamic solution for all $\gpol>1$, then recover it from a singular auxiliary kernel for $1<\gpol\leq3$.

\subsection{Direct hydrodynamic solution}
Consider a polytropic equation of state
\begin{equation}
    \epsilon(\rho) = \frac{A_{\pol}}{\gpol(\gpol - 1)}\rho^\gpol, \qquad A_\pol > 0, \quad \gpol > 1. 
\end{equation}
The case $\gpol=3$ includes the Calogero--Sutherland gas and free fermions, while $\gpol=2$ corresponds to the leading mean-field equation of state of the weakly coupled Bose gas. The sound velocity and acoustic coordinate follow directly from the equation of state:
\begin{equation}
    c_s^2=\rho\,\epsilon''(\rho)
    =A_\pol \rho^{\gpol-1},
    \qquad
    Q=\int_0^\rho \frac{c_s(\rho')}{\rho'}\,d\rho'
    =\frac{2c_s}{\gpol-1}.
\end{equation}
For the hydrodynamic solution, it is convenient to use $d=2/(\gpol-1)>0$. This is related to the parameter $n$ of Refs.~\cite{yeh2022emptiness,abanov2025polytropicEFP} by $d=2n+1$. Then $Q=dc_s$ and $\rho\propto Q^d$. The Luttinger parameter follows a power law $K=\pi\rho/c_s\propto Q^{d-1}$, and the hodograph equation reduces to
\begin{equation}
    \mathcal W_{QQ} + \frac{d - 1}{Q}\mathcal W_Q + \mathcal W_{vv} = 0. 
\end{equation}
This is the radial Laplace equation in $d+1$ dimensions, with radial coordinate $Q$ and an additional coordinate $v$, continued to real $d>0$.

With $\nu = d/2 - 1$, the regular modes normalized by $B_p(0)=1$ are
\begin{equation}
    B_p(Q) = 2^\nu \Gamma(\nu + 1) (pQ)^{- \nu} J_\nu (pQ),
\end{equation}
where $J_\nu$ is the Bessel function of the first kind. Imposing the EFP mixed boundary conditions fixes the boundary values $F(Q)=-\mathcal W(Q,0)$:
\begin{equation}\label{eq:polytropic-FQ}
    F(Q)=\frac{R}{C_d}(q^2-Q^2)_+^{1/2},
    \qquad
    C_d=\frac{\sqrt{\pi}\,\Gamma\left(\frac{d+1}{2}\right)}
    {\Gamma(d/2)},
\end{equation}
Here $x_+=\max(x,0)$, so $F$ vanishes on the Dirichlet segment $Q>q$.

The same modes extend $\mathcal W(Q,0)=-F(Q)$ into $v<0$. Evaluating the resulting potential at $Q=0$ determines the vacuum boundary, as shown in Appendix~\ref{app:polytropic-details}. On the upper-right branch ($x,\tau>0$), where $v<0$, set $\eta=|v|/q$. The characteristic intercept is
\begin{equation}
    X(v)=R+\partial_v\mathcal W(0,v)
        =R\,b_d(\eta),
\end{equation}
with
\begin{equation}
    b_d(\eta) = N_d \int_0^\eta \frac{ds}{(1 + s^2)^{(d+2)/2}}, 
    \qquad 
    N_d = \frac{2\Gamma\left( \frac{d+2}{2} \right)}
    {\sqrt\pi\,\Gamma\left(\frac{d+1}{2} \right)}. 
\end{equation}
The envelope relations $\tau=-X'(v)$ and $x=X(v)-vX'(v)$ give the vacuum boundary as
\begin{equation}
    \frac{x(\eta)}{R} = b_d(\eta) - \eta b_d'(\eta), \quad \frac{q \tau(\eta)}{R} = b_d'(\eta ).
\end{equation}
The action follows from the same boundary values $F$. With $q=dc_{s0}$, the weight is $K(Q)=\pi d\rho_0Q^{d-1}/q^d$, and the boundary integral~\eqref{eq:hydro-action-trace} gives
\begin{equation}
    \mathcal I_{\mathrm{EFP}}  = \frac{2R}{\pi}\int_0^q K(Q)F(Q)\,dQ = \rho_0 c_{s0} R^2 \frac{d^2}{d + 1} \left[ \frac{\Gamma(d/2)}{\Gamma((d+1)/2)} \right]^2,
\end{equation}
This reproduces the polytropic EFP action~\cite{yeh2022emptiness,abanov2025polytropicEFP}.

\subsection{A singular auxiliary kernel}
We now construct an auxiliary kernel that reproduces the same equation of state and recovers the hydrodynamic solution above. For this construction we use $n=(d-1)/2$, which gives a simple form for the kernel operator and the rapidity density below. The kernel family considered here has real $n\geq0$, corresponding to $d\geq1$ or $1<\gpol\leq3$; the direct hydrodynamic solution also covers $0<d<1$.

The vacuum behavior explains why a singular kernel is needed: for $n>0$, the power law $K\propto Q^{2n}$ vanishes at $Q=0$ and cannot satisfy the regular-kernel condition $K(0)=1$. We therefore allow a distributional kernel and define its whole-line operator by the Fourier multiplier
\begin{equation}
    1 - \mathcal T_n = a|\partial_{\lambda}|^{2n}, 
    \quad 
    \widehat \Delta_n (p) = 2\pi\bigl(1 - a |p|^{2n}\bigr), 
    \quad a >0.
\end{equation}
Each function is extended by zero outside its support before applying the full operator, including its contact terms. We first check that this kernel gives the required equation of state. The identity for compactly supported powers in Appendix~\ref{app:polytropic-details} gives the auxiliary rapidity density on $[-Q,Q]$ and its thermodynamic moments:
\begin{equation}
    \rho_p(\lambda;Q) = \frac{(Q^2 - \lambda^2)_+^n}{2\pi a\Gamma(2n + 1)}, 
    \qquad 
    \rho \propto Q^{2n+1}, 
    \quad 
    \epsilon\propto Q^{2n+3}, \quad
    c_s = \frac{Q}{2n+1}. 
\end{equation}
Eliminating $Q$ gives $\epsilon\propto\rho^{\gpol}$, with $\gpol=1+2/(2n+1)$, as required by $d=2n+1$. The parameter $a$ fixes the density scale and is chosen to reproduce $A_\pol$.

With the kernel fixed in this way, the EFP boundary equation~\eqref{eq:chi-compact-equation} becomes
\begin{equation}
    a |\partial_{\lambda}|^{2n+1} \chi_n = R,
    \qquad |\lambda|<q.
\end{equation}
The same power-function identity, now with exponent $n+1/2$, gives $\chi_n$, and differentiation yields $\mathfrak g_n$,
\begin{equation}
    \chi_n = \frac{R(q^2 - \lambda^2)_+^{n + 1/2}}{a\Gamma(2n + 2)},
    \qquad
    \mathfrak g_n = - \frac{2\chi'_n}{\pi R}
    = \frac{2\lambda(q^2 - \lambda^2)^{n - 1/2}}
    {\pi a \Gamma(2n+1)},
\end{equation}
for $|\lambda|<q$. The function $\chi_n$ vanishes at both endpoints, while $\mathfrak g_n$ is odd and integrable for all $n\geq0$.

To compare with the direct solution, we verify the potential and action separately in Appendix~\ref{app:polytropic-details}, without assuming that the regular-kernel endpoint identities apply to this distributional kernel. The Fourier--Bessel transform of $\chi_n$, combined with the modes $B_p(Q)$ at $d=2n+1$, recovers the same hydrodynamic potential and hence the vacuum boundary. The first moment of $\mathfrak g_n$ reproduces the action above. Thus, the auxiliary-kernel construction recovers the polytropic instanton without relying on microscopic integrability.

\begin{figure}[htbp]
    \centering
    \includegraphics[width=\linewidth]{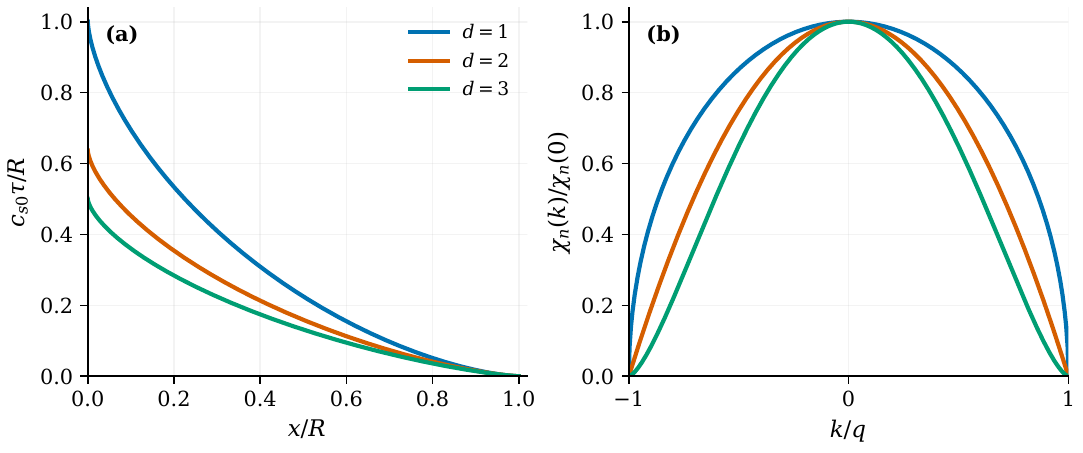}
    \caption{Exact polytropic solutions at $d=1,2,3$, corresponding to $\gpol=3,2,5/3$. (a) Upper-right vacuum boundaries in acoustic coordinates. The first two cases correspond to free fermions and the mean-field Bose gas, respectively. (b) The normalized functions $\chi_n$ for the same three values, with $n=(d-1)/2$.}
    \label{fig:polytropic-examples}
\end{figure}

Figure~\ref{fig:polytropic-examples} illustrates these results for $d=1,2,3$, or $n=0,\tfrac12,1$. In acoustic time, the upper-right vacuum boundary is parametrized by $(x/R,c_{s0}\tau/R)=(b_d-\eta b_d',b_d'/d)$. For $d=1$, this reduces to the free-fermion astroid, $(x/R)^{2/3}+(c_{s0}\tau/R)^{2/3}=1$. The corresponding normalized functions $\chi_n(k)/\chi_n(0)=[1-(k/q)^2]^{d/2}$ illustrate how the behavior near the endpoints depends on the equation of state.

\section{Discussion and conclusions}
For the Lieb--Liniger gas, we have derived from hydrodynamics the integral equation for the previously conjectured leading EFP rate~\cite{khanikati2026efp}. We have also constructed the full instanton, including its density and current profiles and vacuum boundary. KKN's elliptic solution~\cite{kazakov1999dparticles} gives explicit expressions for the rate and vacuum boundary; reconstructing the liquid profile additionally requires Lieb--Liniger thermodynamics. Direct hydrodynamic minimization supports the predicted vacuum geometry, although numerical uncertainties remain near the cusps.

The broader contribution is an auxiliary-kernel method for dispersionless hydrodynamics in one-dimensional Galilean fluids. For an equation of state admitting a suitable kernel representation, the EFP mixed boundary conditions reduce to a single integral equation on a finite interval. Solving this equation, analytically or numerically, determines the action, vacuum boundary, and full instanton through the formulas summarized in Sec.~\ref{sec:practical-procedure}. This construction does not require microscopic integrability, but it relies on the operator and regularity assumptions of Sec.~\ref{sec:auxiliary-kernel}. Singular kernels must be checked separately, as we do for the polytropic family in Sec.~\ref{sec:polytropic}. This kernel construction recovers the known instanton for $1<\gpol\leq3$; the direct hydrodynamic solution applies more broadly, to all $\gpol>1$.

The solutions are exact within dispersionless hydrodynamics under the stated mathematical assumptions. To identify their action with the leading large-$R$ microscopic EFP exponent, the reconstructed configuration must be a globally single-valued minimizing saddle. Gradient and microscopic endpoint corrections must also remain subleading, as assumed in Sec.~\ref{sec:macroscopic-limit}. We do not determine the subleading terms. Within a hydrodynamic approach, estimating endpoint corrections would require matching to a microscopic or gradient-regularized solution near the endpoints, while computing the fluctuation prefactor would require analyzing fluctuations about the saddle.


The spectral modes depend only on the fluid, so they can be reused for other boundary conditions by finding new amplitudes (Sec.~\ref{sec:practical-procedure}). A concrete next application is partial depletion: fix the particle number in an interval to a fraction of its mean and determine the action as a function of that fraction. This requires deriving the corresponding hodograph boundary data and amplitudes, reconstructing the flow, and selecting the minimizing saddle. An extension beyond the Galilean setting, to the XXZ spin chain, will be presented in forthcoming work by C.~Jordan and A.~G.~Abanov~\cite{jordanAbanovXXZ}.

\paragraph*{Note added.}
Shortly before posting this manuscript, we learned of the preprint by Huang and Zhu~\cite{huangzhu2026quantumcost}. Starting from the microscopic Fredholm determinant, they recover the leading Lieb--Liniger emptiness rate conjectured in our earlier work~\cite{khanikati2026efp}. They also derive subleading terms in the large-interval expansion at fixed finite repulsion.
Their electrostatic construction solves the same auxiliary integral equation and gives an elliptic expression for the rate equivalent to the one obtained here from the KKN parametrization. The hydrodynamic construction developed here also gives the spacetime instanton, including its density and current profiles and its vacuum boundary. The hydrodynamic method for computing the leading rate also applies beyond microscopically integrable models when the equation of state admits an auxiliary kernel representation, as illustrated here for a family of polytropic gases.

\section*{Acknowledgements}
We thank John Anderson, Theodore Drivas, Dmitri Gangardt, and Sasha Zamolodchikov for helpful discussions.

We used OpenAI's ChatGPT and Anthropic's Claude to assist with drafting and revising the manuscript, exploring and checking analytical derivations, and developing numerical calculations. The authors take full responsibility for the final manuscript and its scientific claims.


\section*{Data and code availability}
The code and data required to reproduce the main numerical results and figures of this work are available on GitHub~\cite{AbanovKhanikati2026code}.

\appendix

\section{Hydrodynamic action in hodograph variables}\label{app:hodograph-action}
Here, we derive $\mathcal I_{\rm EFP}=E_D$ and the boundary integral for the action~\eqref{eq:hydro-action-trace} by transforming the liquid action and evaluating the boundary and vacuum contributions. We then derive the hodograph equation off shell and show how the contour terms account for the difference between the negative bulk action and the positive on-shell result. This derivation does not use an auxiliary kernel.

We assume an admissible minimizing saddle that can be globally reconstructed. The equality $\mathcal I_{\rm EFP}=E_D$ holds provided the regularity conditions stated below and vanishing endpoint contributions. Expressing the action as a Dirichlet boundary integral also requires finite Dirichlet energy and sufficient decay at hodograph infinity. Appendix~\ref{app:ll-endpoint} verifies the required endpoint conditions for the Lieb--Liniger gas under the stated assumptions, while Appendix~\ref{app:kernel-action} expresses the action as a moment of $\chi$.

Throughout, $m=\hbar=1$ and $v\leq0$ on the upper-right quadrant. Define
\begin{equation}
 P(\rho)=\rho\mu(\rho)-\epsilon(\rho),\qquad P_0=P(\rho_0),\qquad
 A(\rho,v)=\frac{\rho v^2}{2}+U(\rho),
\end{equation}
where $U$ is the background-subtracted energy density. Since $\epsilon(0)=P(0)=0$ and $U(0)=P_0$, the vacuum contributes $P_0$ times its spacetime area.

On a nondegenerate liquid chart, write the inverse map as
\begin{equation*}
 T(Q,v)=\tau(Q,v)=\frac{\mathcal W_Q}{c_s},\qquad
 x(Q,v)=R+\mathcal W_v+vT.
\end{equation*}
Let $D$ be a smooth truncated domain representing liquid in one physical quadrant, with
\begin{equation*}
 J=x_QT_v-x_vT_Q>0,\qquad h(Q)=\rho c_s,
 \qquad \nabla=(\partial_Q,\partial_v).
\end{equation*}
Here $n$ and $d\ell$ denote the outward unit normal and line element in hodograph space. We define the truncated Dirichlet energy by
\begin{equation*}
 E_D(D)=4\int_D\frac{dQ}{2\pi}\,dv\,K|\nabla\mathcal W|^2,
\end{equation*}
where the factor of four accounts for the four spacetime quadrants. Taking the limit as $D$ covers the liquid hodograph domain gives $E_D$. We use the mixed boundary conditions and the finite imaginary-time limit at the vacuum boundary specified in Sec.~\ref{sec:hodograph-problem}.

\subsection{Physical action on the instanton}\label{app:action-onshell}
On a hydrodynamic solution, the physical liquid action is $\mathcal I_L(D)=\int_D AJ\,dQ\,dv$. The inverse Euler equations give
\begin{equation}
 x_Q=vT_Q+c_sT_v,\qquad x_v=vT_v-c_sT_Q,
 \qquad J=c_s|\nabla T|^2.
\end{equation}
Their compatibility condition is $\nabla\cdot(h\nabla T)=0$. Put
\begin{equation}
 f=\frac{A}{\rho}=\frac{v^2}{2}+\frac{U}{\rho}.
\end{equation}
Thermodynamics yields
\begin{equation}
 hf_Q=P-P_0,\qquad hf_v=hv,\qquad \nabla\cdot(h\nabla f)=2h.
\end{equation}
Consequently,
\begin{equation}\label{eq:app-first-onshell-identity}
 hf|\nabla T|^2=hT^2+\nabla\cdot\boldsymbol B,
 \qquad \boldsymbol B=hfT\nabla T-\frac12hT^2\nabla f.
\end{equation}
Using $AJ=hf|\nabla T|^2$ and $hT^2=K\mathcal W_Q^2/\pi$, we express the liquid action as a bulk integral involving $\mathcal W_Q$ and a boundary integral of $\boldsymbol B$. To recover the full Dirichlet energy, we use the hodograph equation $\nabla\cdot(K\nabla\mathcal W)=0$ and the fact that $K$ is independent of $v$. The vector field
\begin{equation}
 \boldsymbol S
 =vK|\nabla\mathcal W|^2\boldsymbol e_v
 -2vK\mathcal W_v\nabla\mathcal W
\end{equation}
then satisfies $\nabla\cdot\boldsymbol S = K(\mathcal W_Q^2-\mathcal W_v^2).$ Thus, on every smooth truncated domain,
\begin{equation}\label{eq:app-onshell-finite}
 \mathcal I_L(D)=\frac14E_D(D)
 +\oint_{\partial D}\left(\boldsymbol B+\frac{\boldsymbol S}{2\pi}\right)\cdot n\,d\ell.
\end{equation}
\subsection{Boundary terms and the optimal action}\label{app:action-boundaries}
Let $\mathcal A_+$ denote the vacuum area in one spacetime quadrant. We now take the limit of Eq.~\eqref{eq:app-onshell-finite} as $D$ covers the full liquid hodograph domain in that quadrant. Adding the vacuum action and using reflection symmetry in $x$ and $\tau$ gives
\begin{equation}\label{eq:app-action-full-contours}
 \mathcal I_{\rm EFP}-E_D
 =4\lim_{D}\oint_{\partial D}
 \left(\boldsymbol B+\frac{\boldsymbol S}{2\pi}\right)\cdot n\,d\ell
 +4P_0\mathcal A_+.
\end{equation}
We keep the contour contributions at the mixed junction and hodograph infinity, evaluating their limits separately when both limits exist.

\paragraph*{Vacuum boundary.}
Parametrize the vacuum boundary by $(x_e(v),T_0(v))$, where $T_0(v)=T(0,v)$. The reconstruction equations give $x_e'=vT_0'$, so
\begin{equation}
 \mathcal A_+=-\int_{-\infty}^0T_0x_e'\,dv
 =\frac12\int_{-\infty}^0T_0^2\,dv
 -\frac12[vT_0^2]_{-\infty}^0.
\end{equation}
We assume that $T$ and $A$ remain finite as $Q\to0$, that $\rho T_\rho\to0$ and $\rho T\mathcal W_v\to0$, and that the corresponding integrals converge. The outward normal is $-\boldsymbol e_Q$, giving 
\begin{equation}
 \int_{\rm vac}\boldsymbol B\cdot n\,d\ell
 =-\frac{P_0}{2}\int_{-\infty}^0T_0^2\,dv,
 \qquad
 \int_{\rm vac}\boldsymbol S\cdot n\,d\ell=0.
\end{equation}
In Eq.~\eqref{eq:app-action-full-contours}, the contribution from $\boldsymbol B$ cancels the vacuum-area term, leaving only the endpoint contribution.

\paragraph*{Boundary at $v=0$ and the mixed junction.} On $v=0$ away from $Q=q$, the flux of $\boldsymbol B$ vanishes. For $Q>q$, we have $T=0$, while for $Q<q$, differentiating $\mathcal W_v(Q,0)=-R$ gives $T_v=0$. The flux of $\boldsymbol S$ also vanishes as it is proportional to $v$.

Near the mixed junction $(Q,v)=(q,0)$, corresponding to physical spacetime infinity, define $r=|(Q-q)+iv|$ and assume
\begin{equation*}
 \mathcal W=O(r^{1/2}),\qquad
 \nabla\mathcal W=O(r^{-1/2}),\qquad
 \nabla^2\mathcal W=O(r^{-3/2}).
\end{equation*}
Background subtraction gives $f=((Q-q)^2+v^2)/2+O(r^3)$. Consequently, both $\boldsymbol B$ and $\boldsymbol S$ remain bounded, and their contour integrals vanish as the contour shrinks to the junction.

\paragraph*{Slit endpoint.} On a large contour $\Gamma_L$ at hodograph infinity, $Q^2+v^2\to\infty$, the remaining flux is
\begin{align}\label{eq:app-slit-contour}
 C_L=\int_{\Gamma_L}\Bigl\{&c_sAT\partial_nT
 -\frac{T^2}{2}\bigl[(P-P_0)n_Q+hv n_v\bigr]\notag\\
 &+\frac{vK}{2\pi}\bigl[|\nabla \mathcal W|^2n_v-2\mathcal W_v\partial_n\mathcal W\bigr]\Bigr\}d\ell.
\end{align}
The contour approaches the physical slit endpoint as $L\to\infty$. The contributions from its intersections with the vacuum boundary are included in this limit. Using the results for the vacuum boundary and mixed junction, we obtain
\begin{equation}\label{eq:app-action-remainder}
 \mathcal I_{\rm EFP}=E_D+4\lim_{L\to\infty}C_L
 -2P_0[vT_0^2]_{-\infty}^0.
\end{equation}
Thus $\mathcal I_{\rm EFP}=E_D$ provided that $C_L\to0$ and $[vT_0^2]_{-\infty}^0=0$.
These conditions do not follow from $\mathcal W\to0$ alone.

\paragraph*{Action as a boundary integral.} Assuming finite Dirichlet energy and a vanishing contribution at hodograph infinity, Green's identity gives
\begin{equation}
 E_D=4\oint\frac{d\ell}{2\pi}\,
 \mathcal W K\partial_n\mathcal W
 =4R\int_0^q\frac{dQ}{2\pi}\,K(Q)F(Q),
 \qquad F(Q)=-\mathcal W(Q,0).
\end{equation}
The contributions from the vacuum boundary and mixed junction vanish, leaving only the Neumann segment $0<Q<q$. Under the stated regularity and endpoint assumptions, the optimal action is therefore
\begin{equation}\label{eq:app-optimal-action}
 \mathcal I_{\rm EFP}=E_D
 =\frac{2R}{\pi}\int_0^q K(Q)F(Q)\,dQ.
\end{equation}

\subsection{Off-shell origin of the hodograph equation}\label{app:action-offshell}
We now derive the hodograph equation before imposing continuity. The transformation
is valid locally wherever the liquid hodograph map is nondegenerate, and its physical interpretation requires the global reconstruction assumed above.

\paragraph*{Constrained action and algebraic elimination.} On a liquid domain $\Omega$, introduce a continuity multiplier $\phi$:
\begin{equation}\label{eq:app-constrained-action}
 S[\rho,j,\phi]=\int_\Omega\left[\frac{j^2}{2\rho}+U(\rho)
 +\phi(\rho_\tau+j_x)\right]dx\,d\tau.
\end{equation}
Integrating by parts, 
\begin{align}\label{eq:app-multiplier-ibp}
 S&=\int_\Omega\left[\frac{j^2}{2\rho}+U(\rho)
 -\rho\phi_\tau-j\phi_x\right]dx\,d\tau+\Lambda_\Omega,\\
 \Lambda_\Omega&=\oint_{\partial\Omega}\phi(\rho n_\tau+j n_x)\,d\ell_{\rm phys}.
 \label{eq:app-multiplier-contour}
\end{align}
Stationarity in the fields $j$ and $\rho$ implies
\begin{equation}\label{eq:app-algebraic-fields}
 j=\rho\phi_x,\qquad
 \mu(\rho)=\mu_0+\phi_\tau+\frac{\phi_x^2}{2}.
\end{equation}
For $\mu'(\rho)>0$, these determine $j$ and $\rho$ locally in terms of $\phi$, giving the reduced bulk action
\begin{equation}\label{eq:app-pressure-action}
 S_{\rm bulk,red}[\phi]=\int_\Omega
 \bigl[P_0-P(\rho(\mu_0+\phi_\tau+\phi_x^2/2))\bigr]dx\,d\tau.
\end{equation}
The contour $\Lambda_\Omega$ is evaluated on the same fields. For variations with compact support, 
\begin{equation}
 \delta S_{\rm bulk,red}
 =\int_\Omega\bigl[\rho_\tau+(\rho\phi_x)_x\bigr]\delta\phi\,dx\,d\tau.
\end{equation}
Varying the remaining independent field $\phi$ gives the continuity equation.

\paragraph*{Legendre transformation.} Assume that the gradient map $(x,\tau)\mapsto(v,b)=(\phi_x,\phi_\tau)$ is locally invertible. Its Legendre transform is
\begin{equation}
 V(v,b)=vx+b\tau-\phi,\qquad dV=x\,dv+\tau\,db.
\end{equation}
Reparametrize the gradient variables by
\begin{equation}
 b=\mu(\rho(Q))-\mu_0-\frac{v^2}{2},\qquad db=c_s(Q)\,dQ-v\,dv,
\end{equation}
since $\mu'(\rho)\rho_Q=c_s$. In these variables, $V(Q,v)=\mathcal W+Rv$, and the inverse map is
\begin{equation}\label{eq:app-offshell-map}
 T=\frac{V_Q}{c_s},\qquad x=V_v+vT,\qquad
 \Phi=vx+bT-V,
\end{equation}
where $x=x(Q,v)$, $T=\tau(Q,v)$, and $\Phi=\phi(x,T)$.

On the orientation-preserving chart $D$, put $H(Q)=P_0-P(\rho(Q))$, so $H'=-h$. The reduced bulk action is $\int_D HJ\,dQ\,dv$. Integrating the Jacobian terms by parts using Eq.~\eqref{eq:app-offshell-map} and $\rho_Q=\rho/c_s=K/\pi$ gives, without imposing continuity,
\begin{equation}\label{eq:app-offshell-local}
 HJ=-\frac{K}{2\pi}|\nabla V|^2+\nabla\cdot\boldsymbol G,
\end{equation}
where
\begin{equation}\label{eq:app-legendre-flux}
 \boldsymbol G=\left(HxT_v+\frac{\rho}{2}V_v^2,
 -HxT_Q+\frac{\rho v}{2c_s}V_Q^2\right).
\end{equation}
The original boundary term $\Lambda_\Omega$ becomes
\begin{equation}\label{eq:app-pulled-multiplier}
 \Lambda_D=\oint_{\partial D}\Phi\rho
 \left[(vT_v-x_v)n_Q+(x_Q-vT_Q)n_v\right]d\ell.
\end{equation}
Combining these expressions, we obtain the off-shell reduced action
\begin{equation}\label{eq:app-offshell-action}
 S_{\rm red}(D)=-\frac1{2\pi}\int_D
 K|\nabla V|^2\,dQ\,dv
 +\oint_{\partial D}\boldsymbol G\cdot n\,d\ell+\Lambda_D.
\end{equation}
Here $V$ is the independent field, while $x$, $T$, and $\Phi$ are determined by Eq.~\eqref{eq:app-offshell-map}. Variations of $V$ with compact support in the interior leave the boundary terms unchanged and preserve local invertibility. Thus,
\begin{equation}
 \delta S_{\rm red}=\frac1\pi\int_D
 \nabla\cdot(K\nabla V)\,\delta V\,dQ\,dv.
\end{equation}
Since $K$ is independent of $v$, subtracting $Rv$ gives the same hodograph equation~\eqref{eq:hodograph-equation} for $\mathcal W$. Its equivalence to continuity follows from
\begin{equation}
 J\bigl[\rho_\tau+(\rho v)_x\bigr]
 =\rho_Q(vT_v-x_v)-\rho T_Q
 =-\frac1\pi\nabla\cdot(K\nabla V).
\end{equation}
The negative bulk term does not define a minimization principle for $V$. Moreover, the boundaries of the hodograph domain may change under variations of the physical fields. Thus, the derivation establishes the hodograph equation, but not the equivalence of the full variational problems.

At fixed $R$, the shift $\mathcal W=V-Rv$ changes the boundary flux to
\begin{equation}
 \boldsymbol G_R=\boldsymbol G
 -\frac{K}{2\pi}(2R\mathcal W+R^2v)\boldsymbol e_v.
\end{equation}
The reduced action can then be written as
\begin{equation}\label{eq:app-offshell-subtracted}
 S_{\rm red}(D)=-\frac14E_D(D)
 +\oint_{\partial D}\boldsymbol G_R\cdot n\,d\ell+\Lambda_D.
\end{equation}
The factor $1/4$ arises because $D$ represents one spacetime quadrant. Both boundary integrals must be included, and the vacuum contribution is added separately.

\paragraph*{Relation to the physical action.} On shell, the continuity equation makes the multiplier term in Eq.~\eqref{eq:app-constrained-action} vanish. Thus, $S_{\rm red}(D)|_{\rm on\ shell} =\mathcal I_L(D)=\int_D AJ\,dQ\,dv$. Equations~\eqref{eq:app-offshell-subtracted} and~\eqref{eq:app-onshell-finite} express this same action with bulk contributions $-E_D(D)/4$ and $+E_D(D)/4$, respectively. The difference is accounted for by the boundary terms:
\begin{equation}\label{eq:app-contour-compensation}
 \oint_{\partial D}\left(
 \boldsymbol G_R-\boldsymbol B
 -\frac{\boldsymbol S}{2\pi}
 \right)\cdot n\,d\ell+\Lambda_D
 =\frac12E_D(D).
\end{equation}
The positive on-shell action therefore results from including the boundary terms, not from changing the sign of the off-shell bulk action.

\section{Auxiliary-kernel method and action}\label{app:kernel-action}
We derive the spectral identities and equivalent boundary equations used in Sec.~\ref{sec:auxiliary-kernel}, then express the boundary integral in Eq.~\eqref{eq:hydro-action-trace} as a moment of $\chi$. The same dressed endpoint identity also yields a quadrature for the Lieb--Liniger liquid profile.

\subsection{Dressed edge identities and the spectral equation}\label{app:dressed-modes}
We collect here the dressed edge identities and the resulting spectral equation, whose ingredients are scattered across the Bethe ansatz and earlier integral-equation literature. Haldane~\cite{Haldane1981BetheAnsatz} and Korepin and Slavnov~\cite{KorepinSlavnov1998Scattering} combined cutoff and rapidity derivatives to differentiate the dressed scattering phase at a moving Fermi edge. We trace the spectral equation Eq.~\eqref{eq:app-dressed-spectral-equation} to Krein's first-order system for finite-interval convolution equations~\cite{Krein1956Accelerants}, developed independently of the Bethe ansatz. Theorem~1.1 of Alpay et al.~\cite{AlpayEtAl2009KreinSystems} gives a statement of Krein's system.

We start from the dressing equation~\eqref{eq:dressing-equation},
\begin{equation*}
 f_Q^{\dr}(\lambda)
 = f(\lambda)+\int_{-Q}^{Q}\frac{d\mu}{2\pi}\,
   \Delta(\lambda-\mu)f_Q^{\dr}(\mu).
\end{equation*}
We hold the bare source $f$ and the real, even difference kernel $\Delta$ fixed as $Q$ varies.
We assume that both the bare source $f$ and the real, even difference kernel $\Delta$ are continuously differentiable on the relevant intervals, that $1-\mathcal T_Q$ is invertible, and that the dressed functions have the derivatives and endpoint values used below. We also assume $Z(Q)=1_Q^{\dr}(Q)\ne0$. These regular-kernel assumptions underlie Sec.~\ref{sec:auxiliary-kernel}. The singular construction in Sec.~\ref{sec:polytropic} is checked separately.

Differentiating Eq.~\eqref{eq:dressing-equation} with respect to $Q$ at fixed $\lambda$ gives
\begin{equation}\label{eq:app-dressing-interval-derivative}
\begin{aligned}
 \partial_Q f_Q^{\dr}(\lambda)
 ={}&\frac{\Delta(\lambda-Q)}{2\pi}f_Q^{\dr}(Q)
 +\frac{\Delta(\lambda+Q)}{2\pi}f_Q^{\dr}(-Q)\\
 &+\int_{-Q}^{Q}\frac{d\mu}{2\pi}\,
   \Delta(\lambda-\mu)\,\partial_Q f_Q^{\dr}(\mu).
\end{aligned}
\end{equation}
This equation also determines the change of the moment. Apply the self-adjoint inverse $(1-\mathcal T_Q)^{-1}$ and integrate over $\lambda$. Transferring the inverse to the constant function gives $1_Q^\dr$, whose dressing equation at either endpoint implies
\begin{equation*}
 \int_{-Q}^{Q}\frac{d\lambda}{2\pi}\,
 \Delta(\lambda\mp Q)1_Q^\dr(\lambda)=Z(Q)-1.
\end{equation*}
The contribution from the change of the dressed function is therefore
\begin{equation*}
 \frac{1}{2\pi}\int_{-Q}^{Q}d\lambda\,\partial_Qf_Q^\dr(\lambda)
 =\frac{Z(Q)-1}{2\pi}\bigl[f_Q^\dr(Q)+f_Q^\dr(-Q)\bigr].
\end{equation*}
Adding the terms from the moving integration limits gives, for any fixed source,
\begin{equation}\label{eq:app-general-moment-derivative}
 \frac{d\mathcal Q[f]}{dQ}
 =\frac{Z(Q)}{2\pi}\bigl[f_Q^\dr(Q)+f_Q^\dr(-Q)\bigr].
\end{equation}
For an even source, the endpoint values are equal and this reduces to Eq.~\eqref{eq:edge-identity}. For an odd source, they cancel: the moment and its derivative both vanish on the symmetric interval. This moment identity is distinct from the derivatives of the dressed endpoint values, which we derive next.

To differentiate with respect to $\lambda$ at fixed $Q$, we use $\partial_\lambda\Delta(\lambda-\mu)=-\partial_\mu\Delta(\lambda-\mu)$ and integrate by parts. This gives
\begin{equation}\label{eq:app-dressing-argument-derivative}
\begin{aligned}
 \partial_\lambda f_Q^{\dr}(\lambda)
 ={}&f'(\lambda)
 -\frac{\Delta(\lambda-Q)}{2\pi}f_Q^{\dr}(Q)
 +\frac{\Delta(\lambda+Q)}{2\pi}f_Q^{\dr}(-Q)\\
 &+\int_{-Q}^{Q}\frac{d\mu}{2\pi}\,
   \Delta(\lambda-\mu)\,\partial_\mu f_Q^{\dr}(\mu).
\end{aligned}
\end{equation}
Adding the two equations cancels the terms from the right endpoint and doubles those from the left endpoint,
\begin{equation*}
\begin{aligned}
 (\partial_Q+\partial_\lambda)f_Q^{\dr}(\lambda)
 ={}&f'(\lambda)+\frac{\Delta(\lambda+Q)}{\pi}f_Q^{\dr}(-Q)\\
 &+\int_{-Q}^{Q}\frac{d\mu}{2\pi}\,
   \Delta(\lambda-\mu)(\partial_Q+\partial_\mu)f_Q^{\dr}(\mu).
\end{aligned}
\end{equation*}
This is a dressing equation with source $f'(\lambda)+\Delta(\lambda+Q)f_Q^{\dr}(-Q)/\pi$. We dress this source on $[-Q,Q]$, keeping the shift in $\Delta(\cdot+Q)$ fixed during dressing. Evaluating the result at $\lambda=Q$ and using the chain rule gives
\begin{equation}
 \frac{d}{dQ}f_Q^{\dr}(Q)
 = (f')_Q^{\dr}(Q)
 +\frac{f_Q^{\dr}(-Q)}{\pi}[\Delta(\cdot+Q)]_Q^{\dr}(Q).
\end{equation}
The dot in \(\Delta(\cdot+Q)\) denotes the rapidity argument being dressed, with \(Q\) held fixed.

Dressing commutes with reflection because the kernel is even and the interval is symmetric. For $f=1$, the dressed values at both endpoints therefore equal $Z(Q)$, and the edge derivative gives
\begin{equation}
 \frac{1}{\pi}[\Delta(\cdot+Q)]_Q^{\dr}(Q)
 =\frac{Z'(Q)}{Z(Q)}.
\end{equation}
Substituting this expression into the edge derivative and differentiating the normalized edge value yields
\begin{equation*}
 \frac{d}{dQ}\frac{f_Q^{\dr}(Q)}{Z(Q)}
 =\frac{(f')_Q^{\dr}(Q)}{Z(Q)}
 +\frac{Z'(Q)}{Z(Q)^2}\bigl[f_Q^{\dr}(-Q)-f_Q^{\dr}(Q)\bigr].
\end{equation*}
For an even source $f$, the bracket vanishes. For an odd source $o$, it equals $-2o_Q^{\dr}(Q)$. With $K(Q)=Z(Q)^2$, the two identities are
\begin{align}
 \frac{d}{dQ}\frac{f_Q^{\dr}(Q)}{Z(Q)}
 &=\frac{(f')_Q^{\dr}(Q)}{Z(Q)},\label{eq:app-even-edge-derivative}\\
 \frac{d}{dQ}\frac{o_Q^{\dr}(Q)}{Z(Q)}
 &=\frac{(o')_Q^{\dr}(Q)}{Z(Q)}
 -\frac{K'}{K}\frac{o_Q^{\dr}(Q)}{Z(Q)}.\label{eq:app-odd-edge-derivative}
\end{align}

Using the moment identity~\eqref{eq:edge-identity}, we obtain $\rho_Q$ and $\mu=\epsilon_Q/\rho_Q$ for the thermodynamics of Sec.~\ref{sec:auxiliary-kernel}. The even edge identity then gives $\mu_Q$,
\begin{equation*}
 \rho_Q=\frac{K}{\pi},\qquad
 \mu=\frac{[\lambda^2/2]_Q^{\dr}(Q)}{Z(Q)},\qquad
 \mu_Q=\frac{[\lambda]_Q^{\dr}(Q)}{Z(Q)}.
\end{equation*}
The odd edge identity yields $\partial_Q(K\mu_Q)=K$. As $Q\to0$ for a regular kernel, $K\mu_Q$ and $\rho$ both vanish. Integration therefore gives $K\mu_Q=\pi\rho$, so that
\begin{equation*}
 c_s^2=\rho\frac{\mu_Q}{\rho_Q}=\left(\frac{\pi\rho}{K}\right)^2,\qquad
 \frac{dQ}{d\rho}=\frac{c_s}{\rho}.
\end{equation*}
As such, $Q$ is the acoustic coordinate with its origin at the vacuum.

Now apply the edge identities to $f(\lambda)=\cos(p\lambda)$ and $o(\lambda)=\sin(p\lambda)$. Their normalized dressed edge values are
\begin{equation*}
 B_p(Q)=\frac{[\cos(p\lambda)]_Q^{\dr}(Q)}{Z(Q)},\qquad
 S_p(Q)=\frac{[\sin(p\lambda)]_Q^{\dr}(Q)}{Z(Q)}.
\end{equation*}
Equations~\eqref{eq:app-even-edge-derivative} and~\eqref{eq:app-odd-edge-derivative} give
\begin{equation}
 B_p'=-pS_p,\qquad S_p'=pB_p-\frac{K'}{K}S_p.
\end{equation}
Differentiating the first relation and using the second gives $B_p''=-p^2B_p-(K'/K)B_p'$. Equivalently,
\begin{equation}\label{eq:app-dressed-spectral-equation}
 \frac{1}{K}\partial_Q(K\partial_QB_p)=-p^2B_p.
\end{equation}
As $Q\to0$, the dressing operation approaches the identity, giving $Z\to1$, $B_p\to1$, and $S_p\to0$. Together with $B_p'(0)=0$ and $K\to1$, these limits establish the vacuum normalization and regular weighted-flux condition of Sec.~\ref{sec:spectral}.

\subsection{Cauchy resolvent and the supported boundary equation}\label{app:resolvent-equivalence}
For the construction of Sec.~\ref{sec:compact-reconstruction}, keep $q$ fixed and take $\chi$ real, even, and absolutely continuous on $[-q,q]$, with $\chi(\pm q)=0$, $\chi'\in L^1$, and zero extension outside. It determines the amplitudes~\eqref{eq:chi-spectral-amplitude} and the Cauchy transform~\eqref{eq:resolvent-definition}, whose large-$z$ expansion is
\begin{equation}\label{eq:hoppe-cauchy-transform}
    C(z)=\frac1{\pi R}\int_{-q}^q\frac{\chi(k)}{z-k}\,dk
    =\frac A z+O(z^{-3}),\qquad
    A=\frac1{\pi R}\int_{-q}^q\chi(k)\,dk.
\end{equation}
Since $\chi$ is real and even, the Cauchy transform has the Schwarz symmetry and oddness described in Sec.~\ref{sec:compact-reconstruction}, and its $z^{-2}$ coefficient vanishes. The endpoint conditions and Cauchy-integral representation exclude the homogeneous term $1/\sqrt{z^2-q^2}$, as well as terms with endpoint poles, such as $z/(z^2-q^2)$. The latter have no jump across the interior of the cut. Both types can be analytic away from the cut and decay at infinity, so the additional conditions are necessary. In particular, absolute continuity of $\chi$ and its vanishing at the endpoints do not guarantee that the Cauchy transform is bounded there.

As before, set $y=-v\geq0$, and define the Hilbert transform and the harmonic extension operator by
\begin{equation*}
    (\mathscr H f)(k)=\frac1\pi\operatorname{PV}\!\int_{\mathbb R}
    \frac{f(\mu)}{k-\mu}\,d\mu,\qquad
    P_y=e^{-y|\partial_k|},\quad y\geq0.
\end{equation*}
The boundary values and upper-half-plane imaginary part are
\begin{equation}\label{eq:app-cauchy-boundary}
    C_\pm=\frac{\mathscr H\chi\mp i\chi}{R},\qquad
    \operatorname{Im}C(k+iy)=-\frac{P_y\chi(k)}{R}.
\end{equation}
We assume that the convolutions, Fourier multipliers, Hilbert transforms, derivatives, and boundary limits used here and in the next subsection are well defined and that the required operations can be interchanged. For integrable kernels, convolution is defined away from the real axis, but additional regularity is needed to take boundary limits. Distributional kernels require a specified Fourier representation. Finite-interval dressing and its $Q$ derivatives satisfy the separate assumptions of Sec.~\ref{app:dressed-modes}, including $Z(Q)\ne0$.

Put
\begin{equation}\label{eq:chi-source}
    H=(1-\mathcal T_\infty)\chi,\qquad
    \supp\chi\subset[-q,q].
\end{equation}
Commuting whole-line convolution with $P_y$ gives
\begin{equation}\label{eq:chi-reconstruction}
    \mathcal W(Q,-y)=-\frac{[P_yH]^\dr_Q(Q)}{Z(Q)},\qquad
    H(k)=\int_0^\infty a(p)\cos(pk)\,dp,
\end{equation}
Here, $a(p)=[1-\kappa(p)]\widehat\chi(p)/\pi$, as in Eq.~\eqref{eq:chi-spectral-amplitude}. Equation~\eqref{eq:chi-reconstruction} reproduces the spectral formula~\eqref{eq:chi-spectral-reconstruction}, while Eq.~\eqref{eq:app-cauchy-boundary} gives the resolvent formula~\eqref{eq:resolvent-reconstruction}. For $Q\geq q$, extending $\chi$ by zero gives $H|_{[-Q,Q]}=(1-\mathcal T_Q)\chi$. Its dressed edge value therefore vanishes for $Q>q$, satisfying the exterior boundary condition.

Taking the real part of Eq.~\eqref{eq:app-cauchy-boundary} converts the resolvent equation to $\mathscr H H=Rk$ on $(-q,q)$. Using $\partial_k\mathscr H=|\partial_k|$,
we differentiate to obtain
\begin{equation}\label{eq:chi-compact-equation}
    \mathcal D_\Delta\chi=R\quad(|k|<q),\qquad
    \mathcal D_\Delta=|\partial_k|(1-\mathcal T_\infty),\qquad
    \chi=0\quad(|k|>q).
\end{equation}
Conversely, integrating gives $\mathscr H H=Rk+\mathrm{const}$, where the constant vanishes by oddness. Thus the resolvent equation and the supported boundary equation are equivalent under the stated assumptions.

To prove that the resolvent equation is necessary for the Neumann boundary condition, define $e(k)=\operatorname{Re}C'_{\Delta,+}(k)-1$ and $u_Q=(1-\mathcal T_Q)^{-1}e$ for $0<Q<q$. The even function $e$ is independent of $Q$ and is assumed to be continuously differentiable on each interval, as in Sec.~\ref{app:dressed-modes}. Equation~\eqref{eq:resolvent-neumann} gives $u_Q(\pm Q)=0$, and Eq.~\eqref{eq:app-dressing-interval-derivative} then gives $\partial_Q u_Q(k)=0$ at fixed $|k|<Q$. Taking $Q\to|k|^+$ gives $u_Q(k)=0$, with continuity extending the result to $k=0$. The dressing equation then implies $e=0$. This argument requires regularity of the dressed functions and their endpoint values, as well as invertibility, and does not directly apply to singular kernels. 

The function $\chi$ can also be reconstructed from an integrable odd function $\mathfrak g$:
\begin{equation}\label{eq:chi-from-g}
    \chi(k)=\frac{\pi R}{2}\int_k^q\mathfrak g(\mu)\,d\mu,
    \qquad |k|\leq q,
\end{equation}
extended by zero outside $[-q,q]$. Since $\mathfrak g$ is odd, $\chi$ is even and vanishes at both endpoints, as required for the integral equation and action formula below.

\subsection{Integral equation and electrostatic interpretation}\label{app:integrated-equation}
To obtain the real integral equation~\eqref{eq:mathfrakg} from Eq.~\eqref{eq:chi-compact-equation}, we use the endpoint conditions and operator assumptions of Sec.~\ref{app:resolvent-equivalence}. Since $\chi(\pm q)=0$, differentiating its zero extension introduces no delta functions at the endpoints. The definition~\eqref{eq:mathfrak-g-def} then gives
\begin{equation}\label{eq:app-g-fourier}
    \widehat{\mathfrak g}(p)
    =-\frac{2ip}{\pi R}\widehat\chi(p).
\end{equation}
The assumptions for whole-line operators remain distinct from those for finite-interval dressing in Sec.~\ref{app:dressed-modes}.

\paragraph*{Integrating the operator equation.} The multiplier identity
\begin{equation*}
    |\partial_k|=-\partial_k|\partial_k|^{-1}\partial_k
\end{equation*}
follows from $-(ip)|p|^{-1}(ip)=|p|$ for $p\ne0$. We use the inverse operator only where it is well defined. Since Eq.~\eqref{eq:chi-compact-equation} holds only for $|k|<q$, it cannot be solved by simply dividing by its Fourier multiplier on the whole line. Commuting differentiation with convolution gives
\begin{equation}\label{eq:app-integrated-operator}
    \mathcal D_\Delta\chi
    =\frac{\pi R}{2}\partial_k
      \left[|\partial_k|^{-1}(1-\mathcal T_\infty)\mathfrak g\right]
    =\frac R4\partial_k(L*\mathfrak g).
\end{equation}
The bracketed operator has multiplier $[1-\kappa(p)]/|p|$, giving Eq.~\eqref{eq:integrated-kernel}. Using the definition of $\mathfrak g$, one derivative is absorbed into $\mathfrak g$, and we can integrate the remaining derivative.

Equation~\eqref{eq:chi-compact-equation} gives $L*\mathfrak g=4k+\mathrm{const}$ on $(-q,q)$. Since $L$ is even and $\mathfrak g$ is odd, the constant vanishes, giving Eq.~\eqref{eq:mathfrakg}. An additive constant in $L$ has no effect because $\int_{-q}^q\mathfrak g=0$. Conversely, an integrable odd solution of Eq.~\eqref{eq:mathfrakg} determines an even, absolutely continuous function $\chi$ with $\chi(\pm q)=0$ through Eq.~\eqref{eq:chi-from-g}. If $\chi$ satisfies the stated operator assumptions, Eq.~\eqref{eq:app-integrated-operator} recovers the original equation. This equivalence does not, however, establish existence or uniqueness of solutions.

\paragraph*{Logarithmic potential.} In the Cauchy formulation, $\chi$ represents a line charge in a quadratic external potential. Rewriting the equation in terms of $H'$, where $H=(1-\mathcal T_\infty)\chi$, gives a polarization problem in a uniform field. Assuming the principal values and boundary terms are well defined, we obtain
\begin{equation}\label{eq:app-hypersingular-equation}
    \begin{aligned}
    \frac1\pi\operatorname{PV}\!\int_{\mathbb R}
    \frac{H(k)-H(\mu)}{(k-\mu)^2}\,d\mu
    &=\frac1\pi\operatorname{PV}\!\int_{\mathbb R}
    \frac{H'(\mu)}{k-\mu}\,d\mu\\
    &=R,\qquad |k|<q.
    \end{aligned}
\end{equation}
The first equality follows from integrating by parts, while the second from $|\partial_k|H=R$. Sufficient conditions are that $H$ is locally absolutely continuous on $\mathbb R$, $H\in C^{1,\alpha}_{\rm loc}(-q,q)$ for some $\alpha>0$, $H(\mu)/\mu\to0$ at infinity, and $\int_{\mathbb R}|H'(\mu)|\log(2+|\mu|)\,d\mu<\infty$. Integrating with respect to $k$ gives
\begin{equation}\label{eq:app-logarithmic-equation}
    \frac1\pi\int_{\mathbb R}H'(\mu)
    \log\frac{|k-\mu|}{k_*}\,d\mu=Rk,
    \qquad |k|<q,
\end{equation}
where $k_*>0$ is an arbitrary reference scale. Since $H'$ is odd, the integration constant vanishes, and the result is independent of $k_*$. For singular kernels, these regularity conditions need not hold, and the operator derivation instead requires the Fourier-domain assumptions stated above.

Equation~\eqref{eq:app-logarithmic-equation} describes a signed charge density proportional to $H'$ whose logarithmic potential cancels the external potential $-Rk$ on $(-q,q)$. The applied field is uniform, and the total charge vanishes since $H'$ is odd and integrable. The relation
\begin{equation*}
    H'=\chi'-\frac{\Delta}{2\pi}*\chi'
\end{equation*}
shows why this charge generally extends beyond $[-q,q]$. Although $\chi'$ is supported on this interval, its convolution with the kernel need not be. Therefore, the integrals in Eqs.~\eqref{eq:app-hypersingular-equation} and~\eqref{eq:app-logarithmic-equation} must extend over the whole line. For $\Delta=0$, the problem reduces to a conducting segment in a uniform field.

Substituting $H'=(1-\mathcal T_\infty)\chi'$ and $\chi'=-\pi R\mathfrak g/2$ into Eq.~\eqref{eq:app-logarithmic-equation} gives the kernel directly,
\begin{equation}\label{eq:app-logarithmic-kernel}
    L(x)=-2\log\frac{|x|}{k_*}
    +\frac1\pi\int_{\mathbb R}\Delta(t)
       \log\frac{|x-t|}{k_*}\,dt.
\end{equation}
This expression is valid when the integrals converge and their orders can be interchanged. For a regular kernel, it is sufficient that $\int|\Delta(t)|\log(2+|t|)\,dt<\infty$. Its Fourier transform gives Eq.~\eqref{eq:integrated-kernel} for $p\neq 0$, up to irrelevant additive constants in $L$. When the real-space integrals are not defined, $L$ is instead specified by its Fourier multiplier, in the distributional sense if necessary.

\subsection{Action from boundary values}\label{app:action-boundary-values}
We express the boundary integral~\eqref{eq:hydro-action-trace} as a moment of $\chi$ using Eq.~\eqref{eq:app-general-moment-derivative}. Its derivation differentiates only the dressing interval, so it applies to fixed continuous sources under the invertibility, differentiability, and endpoint assumptions of Sec.~\ref{app:dressed-modes}, with $Z(Q)\ne0$. The $C^1$ source assumption is needed only for the rapidity-derivative identities there.

More generally, for a fixed real, even kernel and fixed continuous even sources $f_1,f_2$, differentiating the dressing equation and using self-adjointness gives
\begin{equation}\label{eq:app-resolvent-derivative}
 \frac{d}{dQ}\langle f_1,(1-\mathcal T_Q)^{-1}f_2\rangle_Q
 =2f_{1,Q}^{\dr}(Q)f_{2,Q}^{\dr}(Q),
\end{equation}
where $\langle a,b\rangle_Q=\int_{-Q}^{Q}a(k)b(k)\,dk$. The factor of $2$ comes from the two endpoints; dependence of the kernel or sources on $Q$ would add terms.

Let $H=(1-\mathcal T_\infty)\chi$, with $\Delta$, $q$, and $\chi$ fixed and $\chi$ extended by zero. Then $H$ is independent of $Q$. By self-adjointness, $\int_{-Q}^Q H_Q^\dr(k)\,dk=2\pi\mathcal Q[H]$. Applying the moment identity to $H$ and using $F=H_Q^\dr(Q)/Z$ from Eq.~\eqref{eq:chi-reconstruction}, together with $K=Z^2$, gives
\begin{equation*}
 \frac{d}{dQ}\int_{-Q}^{Q}H_Q^{\dr}(k)\,dk
 =2Z(Q)H_Q^{\dr}(Q)=2K(Q)F(Q).
\end{equation*}
We assume that this integral vanishes as $Q\to0^+$ and is continuous as $Q\to q^-$. The first limit follows from regular dressing and bounded $H$ near $Q=0$. At $Q=q$, zero extension gives $H|_{[-q,q]}=(1-\mathcal T_q)\chi$, so dressing recovers $\chi$. These limits allow an integrable singularity in $\chi'$ and do not require $C^1$ sources at $k=\pm q$. Integration gives
\begin{equation}
 2\int_0^q K(Q)F(Q)\,dQ=\int_{-q}^q\chi(k)\,dk.
\end{equation}
The action formula~\eqref{eq:hydro-action-trace}, the coefficient $A$ in Eq.~\eqref{eq:hoppe-cauchy-transform}, and integration by parts with $\mathfrak g=-2\chi'/(\pi R)$ and $\chi(\pm q)=0$ give
\begin{equation}\label{eq:app-response-optimal-action}
 \mathcal I_{\rm EFP}=\frac{R}{\pi}\int_{-q}^q\chi(k)\,dk
 =R^2A=\frac{R^2}{2}\int_{-q}^q k\mathfrak g(k)\,dk.
\end{equation}
This derivation applies to regular kernels; the polytropic action is checked separately by direct hydrodynamics.

\subsection{Liquid-profile quadrature from Lieb--Liniger thermodynamics}\label{app:hoppe-profile-quadrature}
The same moment identity gives the liquid-profile quadrature~\eqref{eq:ll-profile-quadrature}. Fix $q,c$ and $y>0$. For the background resolvent $C$, obtained numerically or from the elliptic solution, define
\begin{equation}\label{eq:hoppe-profile-source}
    s_y(k)=\operatorname{Im}\!\left[C(k+iy)-C(k+i(y+c))\right],
    \qquad
    u_Q=(1-\mathcal T_Q)^{-1}s_y.
\end{equation}
The source is real, even, smooth, and independent of $Q$. We retain the invertibility, endpoint regularity, and dressing differentiability assumptions of Sec.~\ref{app:dressed-modes}, with $Z(Q)\ne0$.

The Lieb--Liniger thermodynamic distribution satisfies
\begin{equation}\label{eq:hoppe-profile-thermodynamics}
    (1-\mathcal T_Q)\rho_p(\cdot;Q)=\frac1{2\pi},\qquad
    \rho(Q)=\int_{-Q}^Q\rho_p(k;Q)\,dk,\qquad
    \rho'(Q)=\frac{Z(Q)^2}{\pi}.
\end{equation}
Self-adjointness identifies the moment $J=\mathcal Q[s_y]$ as
\begin{equation}\label{eq:hoppe-profile-pairing}
    J(Q,y)\equiv\int_{-Q}^Q\rho_p(k;Q)s_y(k)\,dk
    =\frac1{2\pi}\int_{-Q}^Q u_Q(k)\,dk.
\end{equation}
Equation~\eqref{eq:app-general-moment-derivative} accounts for both the moving endpoints and the change of dressing, giving
\begin{equation}\label{eq:hoppe-profile-quadrature-derivation}
    \partial_QJ=\frac{Z(Q)u_Q(Q)}{\pi},\qquad
    \frac{R\,\partial_QJ}{\rho'(Q)}
    =R\frac{u_Q(Q)}{Z(Q)}=\mathcal W(Q,-y).
\end{equation}
This proves Eq.~\eqref{eq:ll-profile-quadrature}. The derivative holds $q,c,y,C$ fixed; varying $C$ or $c$ with $Q$ would add terms. The limit $y\to0^+$ requires the boundary regularity of Sec.~\ref{app:resolvent-equivalence}, particularly near the mixed junction.

As $Q\to0^+$, regular dressing gives $\rho_p\to1/(2\pi)$ and $\rho'\to1/\pi$, recovering $\mathcal W(0,-y)=R s_y(0)$. For free fermions, dressing is the identity and evenness gives $\mathcal W(Q,-y)=R s_y(Q)$ for every $Q$.

The quadrature avoids dressing a new source for each $y$ but still requires the Lieb--Liniger thermodynamic distributions. It applies to numerical or elliptic boundary data; it does not give a closed elliptic density formula or rule out one.

\subsection{Existence and uniqueness qualifications}\label{app:compact-uniqueness}
Given a sufficiently regular solution $\chi$, the reconstruction of Sec.~\ref{sec:compact-reconstruction} satisfies the hodograph equation and its boundary conditions at finite $Q$ and $v$. A solution of the full mixed boundary problem must also decay at hodograph infinity, as discussed in Appendix~\ref{app:endpoint-estimates}. The uniqueness of $\chi$ and of the hodograph potential are separate questions. 

For the equation for $\chi$, let the whole-line convolution be real and symmetric, with $1-\kappa(p)\geq0$. On a finite-energy domain where the relevant integrals are defined, the compact equation is the Euler equation of
\begin{equation}
    \frac12\langle \chi,\mathcal D_\Delta\chi\rangle
    -R\int_{-q}^{q}\chi(k)\,dk,
    \qquad
    \mathcal D_\Delta=|\partial_k|(1-\mathcal T_\infty),
\end{equation}
where $\chi$ is extended by zero outside $[-q,q]$. The inner product is taken over the whole line, or equivalently over the support of $\chi$. In the Fourier convention $\widehat\chi(p)=\int e^{-ipk}\chi(k)\,dk$,
\begin{equation}
 \langle\chi,\mathcal D_\Delta\chi\rangle
 =\int_{\mathbb R}\frac{dp}{2\pi}\,|p|\,[1-\kappa(p)]|\widehat\chi(p)|^2.
\end{equation}
If $1-\kappa(p)>0$ for $p\ne0$, the quadratic form is strictly positive for any nonzero $\chi$ supported on $[-q,q]$. The possible zero mode at $p=0$ does not affect this result, since a nonzero constant cannot be supported on a finite interval. Now suppose there are two solutions of Eq.~\eqref{eq:chi-compact-equation}. Their difference satisfies the homogeneous equation, and positivity implies that this difference must vanish. Thus, the finite-energy solution, if it exists, is unique.

Since the equation is invariant under $k\to-k$, the reflection of any solution is also a solution. Uniqueness therefore implies that $\chi$ is even. Even if uniqueness is not established, averaging a solution with its reflection gives an even solution, as used in Sec.~\ref{sec:compact-reconstruction}.

To establish existence, one would also need bounds on the source term and the solution, together with sufficient regularity to reconstruct the instanton and satisfy the endpoint conditions. These  properties do not follow from positivity alone. If $1-\kappa(p)$ changes sign or has additional zeros, the preceding uniqueness argument no longer applies. However, additional zeros do not necessarily imply that solutions are nonunique.

To prove uniqueness of the hodograph potential, suppose that two solutions satisfy the same mixed boundary conditions. Their difference $u$ obeys the homogeneous hodograph equation. Since $K>0$ in the liquid, Green's identity gives $\int K|\nabla u|^2=0$, provided the boundary contributions vanish. It follows that $u$ is constant, with its value fixed by the Dirichlet condition or decay at infinity. Thus, the potential is unique among solutions satisfying these conditions. This does not establish global reconstruction or show that the resulting instanton minimizes the physical action.

\section{Endpoint estimates}\label{app:ll-endpoint}\label{app:endpoint-estimates}
The hydrodynamic action must converge near the singular slit endpoints for the approach to be self-consistent. Their microscopic neighborhoods must also contribute less than the leading macroscopic action. We prove convergence and estimate the omitted hydrodynamic action at a finite cutoff. Estimating its microscopic replacement requires a separate matching calculation.

\begin{figure}[htbp]
    \centering
    \includegraphics[width=0.85\linewidth]{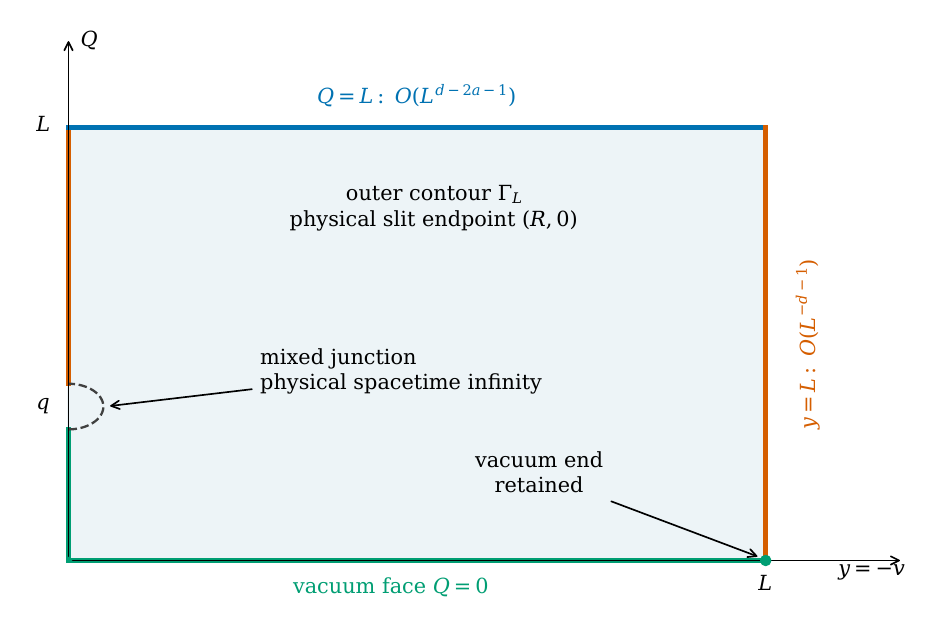}
    \caption{Endpoint contours in $(y,Q)$ coordinates. The two outer sides form $\Gamma_L$ and include its vacuum end at the lower-right corner. The indentation at the mixed junction is treated separately in Appendix~\ref{app:action-boundaries}. The figure is schematic, with $q$ fixed as $L\to\infty$.}
    \label{fig:endpoint-contour}
\end{figure}

The remaining contour in Appendix~\ref{app:hodograph-action} approaches $(x,\tau)=(R,0)$, or $(-R,0)$ on the reflected branch. These points lie at infinity in the hodograph coordinates $(Q,v)$. Put $y=-v\geq0$. Figure~\ref{fig:endpoint-contour} shows the outer contour, including its vacuum end at $Q=0$. The indentation at $(Q,y)=(q,0)$ represents infinity in physical spacetime and was treated in Appendix~\ref{app:action-boundaries}.

\subsection{Power-law weights and assumptions}
Keep the background $q<\infty$ fixed and suppose that
\begin{equation}\label{eq:app-endpoint-coefficients}
 K(Q)\sim k_\infty Q^{d-1},\qquad k_\infty>0,\qquad d>0.
\end{equation}
This describes an asymptotically polytropic equation of state, with energy density and pressure both scaling as $\rho^\gpol$. Free fermions have $d=1$, and finite-coupling Lieb--Liniger has $d=2$ at high density. The thermodynamic identities give    
\begin{equation}
 \rho(Q)\sim\frac{k_\infty}{\pi d}Q^d,
 \qquad c_s(Q)\sim\frac Qd,
 \qquad \gpol=1+\frac2d.
\end{equation}
For an exact power law these relations hold throughout the liquid. Otherwise, they describe only the high-density regime, and derivative estimates require additional control of the asymptotic remainders.

The limiting hodograph operator is
\begin{equation}
 \mathscr P_d=\partial_Q^2+\frac{d-1}{Q}\partial_Q+\partial_y^2.
\end{equation}
A decaying dipole solution, compatible with the homogeneous measurement-line data, is
\begin{equation}\label{eq:app-polytropic-dipole}
 \mathcal W_{\rm dip}=D\frac{y}{(Q^2+y^2)^{(d+1)/2}}.
\end{equation}
This solution is homogeneous of degree $-d$. However, the high-density equation does not establish that this asymptotic form remains valid uniformly near the vacuum boundary. To prove convergence, we therefore use weaker bounds. Assume $K>0$ in the liquid and
\begin{equation}\label{eq:app-general-coefficient-regularity}
 Q\frac{K'}K\longrightarrow d-1,\qquad
 c_s\asymp Q,\qquad |c_s'|\leq C\quad(Q\to\infty).
\end{equation}
To control second derivatives, we also assume that the $C^{0,\alpha}([1/2,2])$ norms of $b_L(s)=LK'(Ls)/K(Ls)$ are uniformly bounded for some $0<\alpha<1$. We require the solution to satisfy the mixed boundary conditions, to decay uniformly as $Q^2+y^2\to\infty$, and to obey
\begin{equation}\label{eq:app-general-large-y}
 |\partial_y^j \mathcal W(Q,-y)|\leq C_j y^{-d-j},
 \qquad j=0,1,2,3,\quad Q\geq0,
\end{equation}
uniformly in $Q$ for sufficiently large $y$. This last bound is an additional assumption, verified below for the polytropic gas and Lieb--Liniger.

We use the assumptions of Appendix~\ref{app:hodograph-action} concerning the vacuum boundary, the differentiated mixed conditions at the junction, and the minimizing reconstruction, together with its notation $A,h,T$. In particular, $K\mathcal W_Q\to0$ at the vacuum and the closing time is finite. Near the vacuum, we assume integrable control of the boundary limits, finite integrals of $K$ and $h$, and $c_s\leq CQ$ on finite intervals. Exact polytropes satisfy these integrability conditions for every $d>0$, even when $K$ diverges. At large $Q$, thermodynamics gives $h=O(Q^{d+1})$ and $|P-P_0|+|U|=O(1+Q^{d+2})$. Away from the junction, the solution remains regular down to $y=0$, so $\mathcal W(Q_*,-y)=O(y)$ for fixed $Q_*>q$.

\subsection{A general exterior barrier}
Choose an exponent $a$ satisfying
\begin{equation}\label{eq:app-barrier-exponent}
 \max\left(0,\frac{d-1}{2}\right)<a<d.
\end{equation}
Such a choice is possible for every $d>0$. The upper bound ensures that the function 
\begin{equation}\label{eq:app-general-barrier}
 w_a(Q,y)=\frac{y}{(Q^2+y^2)^{(a+1)/2}},\qquad
 \mathscr P_K w_a=(a+1)\left(a-1-Q\frac{K'}K\right)
 \frac{y}{(Q^2+y^2)^{(a+3)/2}},
\end{equation}
where $\mathscr P_K=\partial_Q^2+(K'/K)\partial_Q+\partial_y^2$, is a supersolution at large $Q$. The lower bound ensures convergence of the action flux. Since the coefficient in parentheses tends to $a-d<0$, we may choose $Q_*>q$ such that $\mathscr P_K w_a\leq0$ for $Q\geq Q_*$. 

Both functions vanish at $y=0$. On $Q=Q_*$, the bounds $\mathcal W=O(y)$ as $y\to0$ and $\mathcal W=O(y^{-d})$ as $y\to\infty$ imply $|\mathcal W|\leq Cw_a$, since $a<d$. The maximum principle, together with uniform decay at infinity, then gives 
\begin{equation}\label{eq:app-general-exterior-bound}
 |\mathcal W(Q,-y)|\leq C\frac{y}{(Q^2+y^2)^{(a+1)/2}},
 \qquad Q\geq Q_*,\quad y\geq0.
\end{equation}
Near $Q=L$, rescaling balls of radius proportional to $L$ and applying elliptic derivative estimates to the $O(L^{-a})$ bound gives estimates for the first and second derivatives. Odd reflection across $y=0$ extends these estimates to the end of the contour. Using $T=\mathcal W_Q/c_s$, together with $c_s\asymp L$ and bounded $c_s'$, we obtain
\begin{equation}\label{eq:app-general-high-Q}
 \begin{gathered}
 |\mathcal W|\leq CL^{-a},\quad |\nabla \mathcal W|\leq CL^{-a-1},\quad
 |\nabla^2\mathcal W|\leq CL^{-a-2},\\
 |T|\leq CL^{-a-2},\quad |\nabla T|\leq CL^{-a-3},
 \qquad Q=L,\quad 0\leq y\leq L.
 \end{gathered}
\end{equation}

\subsection{Convergence, including the vacuum end}
To control the vacuum end, integrate $(K\mathcal W_Q)_Q=-K\mathcal W_{yy}$ from $Q=0$. Using the vanishing weighted flux at the vacuum and $\int_0^Q K=\pi\rho$, we obtain
\begin{equation}\label{eq:app-general-time}
 T(Q,-y)=-\frac1{\pi\rho(Q)}\int_0^QK(s)\mathcal W_{yy}(s,-y)\,ds,
 \qquad |T|\leq Cy^{-d-2},\quad |T_y|\leq Cy^{-d-3}.
\end{equation}
The normalized integral preserves these bounds uniformly down to $Q=0$, where $T$ has a  continuous limit. Together with $|\mathcal W_Q|=c_s|T|$, this gives
\begin{equation}
 yT_0(-y)^2=O(y^{-2d-3})\longrightarrow0.
\end{equation}

On $Q=L$, we have $A=O(L^{d+2})$, $h=O(L^{d+1})$, and $K=O(L^{d-1})$. Substituting the estimates \eqref{eq:app-general-high-Q} into the flux \eqref{eq:app-slit-contour} gives $O(L^{d-2a-2})$ pointwise. Integrating along the contour then gives $O(L^{d-2a-1})$.

On $y=L$, use Eq.~\eqref{eq:app-general-time}, $|\mathcal W_y|=O(L^{-d-1})$, and $\mathcal W_Q=c_sT$. The thermodynamic growth estimates and vacuum integrability assumptions then bound the three terms in the normal flux by
\begin{equation}
 \int_0^L\!\left[c_s|A T T_y|+LhT^2
 +LK(\mathcal W_Q^2+\mathcal W_y^2)\right]dQ
 =O(L^{-d-1}),\qquad y=L.
\end{equation}
This bound includes the vacuum end $Q=0$. Combining the two contour sides gives
\begin{equation}\label{eq:app-general-contour-vanishing}
 |C_L|\leq C_a L^{d-2a-1}+C L^{-d-1}\longrightarrow0.
\end{equation}
Both powers are negative by Eq.~\eqref{eq:app-barrier-exponent}. The same estimates control the Green contour and the Dirichlet energy in dyadic shells, giving $O(L^{d-2a-1})$ for $Q\geq y$ and $O(L^{-d-1})$ for $y\geq Q$. The sum over these shells therefore converges. The vacuum boundary term also vanishes, since $yT_0(-y)^2\to0$ and $T_0(0)$ is finite. Under the local and physical assumptions stated above, we obtain
\begin{equation}\label{eq:app-ll-action-justified}
 \mathcal I_{\rm EFP}=E_D.
\end{equation}
The boundary integral for the action then follows from Appendix~\ref{app:hodograph-action}. Expressing the action as a moment additionally requires the auxiliary-kernel assumptions of Appendix~\ref{app:kernel-action}.

\subsection{Exact polytropes and free fermions}
For the polytropic equation of state, the Bessel solution of Sec.~\ref{sec:polytropic} and Appendix~\ref{app:polytropic-details} has the dipole~\eqref{eq:app-polytropic-dipole} as its leading term, with corrections of degree at most $-d-2$. The coefficients in its radial expansion are moments of the compactly supported boundary values at $v=0$, weighted by $s^{d-1}$. These moments are finite for every $d>0$, including $0<d<1$, and the expansion in $q/\sqrt{Q^2+y^2}$ can be differentiated uniformly for $\sqrt{Q^2+y^2}>2q$. This establishes Eq.~\eqref{eq:app-general-large-y} and improves the estimates \eqref{eq:app-general-high-Q} to $a=d$. Both contour contributions are then $O(L^{-d-1})$. The potential at $Q=0$, given in Eq.~\eqref{eq:polytropic-wvac}, yields
\begin{equation}
 \mathcal W(0,-y)=D y^{-d}+O(y^{-d-2}),\qquad
 D=-\frac{2R}{\pi}\int_0^q s^{d-1}\sqrt{q^2-s^2}\,ds.
\end{equation}
The regular radial equation gives $T_0=-\mathcal W_{yy}(0,-y)$, confirming the vacuum-tail estimate. This follows directly from the polytropic solution for all $d>0$~\cite{abanov2025polytropicEFP}, without the use of an auxiliary kernel.

For free fermions, $d=1$ and $K=1$, and
\begin{equation}
 \mathcal W(0,-y)=-R\bigl(\sqrt{q^2+y^2}-y\bigr)
 =-\frac{Rq^2}{2y}+O(y^{-3}).
\end{equation}
The harmonic extension satisfies the required derivative bounds, and the endpoint contour contributes $O(L^{-2})$. Thus, the $y^{-1}$ decay at the vacuum is sufficient for convergence of the action.

\subsection{Verification for Lieb--Liniger}\label{app:ll-endpoint-verification}
Fix $0<c<\infty$ and $q<\infty$, and assume the local regularity and uniform hodograph decay specified above, with $\chi\in L^1([-q,q])$. Constants may depend on $c$ and $q$, so the estimates are not uniform in the coupling. The thermodynamic input is
\begin{align}\label{eq:app-ll-endpoint-eos}
 \rho(Q)&\sim\frac{Q^2}{4c},& c_s(Q)&\sim\frac Q2,&
 K(Q)&\sim\frac{\pi Q}{2c},& \frac{K'}K&=\frac1Q+o(Q^{-1}).
\end{align}
These are the weak-coupling thermodynamic asymptotics at high density~\cite{Tracy2016,Prolhac2017}. We also assume sufficient control of the derivatives of the asymptotic remainders to obtain the stated estimate for $K'/K$, bounded $c_s'$, and the scaled $C^{0,\alpha}$ bound. These properties are not proved here from the Bethe equations. At small $Q$, we have $K\to1$, $c_s\sim Q$, and $\rho\sim Q/\pi$. Together with smooth control on finite intervals, these limits give the required vacuum and large-$Q$ bounds.

We now prove the bound
\begin{equation}\label{eq:app-ll-uniform-y-derivatives}
 |\partial_y^m \mathcal W(Q,-y)|\leq C_m y^{-m-2},
 \qquad m=0,1,2,3,
\end{equation}
for all $Q\geq0$ and sufficiently large $y$. Define the Poisson operator by $P_yf=p_y*f$, where $p_y(k)=y/[\pi(k^2+y^2)]$. For Lieb--Liniger, $\mathcal T_\infty=P_c$, so the reconstruction gives 
\begin{equation}\label{eq:app-ll-poisson-difference}
 \mathcal W(Q,-y)=-\frac{[(P_y-P_{y+c})\chi]^\dr_Q(Q)}{Z(Q)}.
\end{equation}
On every finite interval, the positive Lieb--Liniger integral operator has $L^\infty$ norm strictly less than one. Its inverse is therefore positive, and comparison with the dressed identity gives
\begin{equation}\label{eq:app-ll-normalized-dressing-bound}
 \frac{|f_Q^\dr(Q)|}{Z(Q)}
 \leq\|f\|_{L^\infty(\mathbb R)}.
\end{equation}
Importantly, this bound is independent of $Q$. Since $p_y-p_{y+c}=-\int_0^c\partial_y p_{y+s}\,ds$, we obtain
\begin{equation}
 \|\partial_y^m(p_y-p_{y+c})*\chi\|_\infty
 \leq C_m c\|\chi\|_1\,y^{-m-2},
 \qquad y\geq y_*>0.
\end{equation}
Dressing commutes with differentiation in $y$, so Eq.~\eqref{eq:app-ll-normalized-dressing-bound} establishes the required large-velocity bound with $d=2$, uniformly down to $Q=0$.

The general convergence argument now applies, provided the remaining assumptions on the coefficients, uniform decay, and local regularity hold. Choosing $a=2-\varepsilon$, with $0<\varepsilon<3/2$, gives 
\begin{equation}\label{eq:app-ll-endpoint-vanishing}
 |C_L|\leq C_\varepsilon L^{-3+2\varepsilon}
          +CL^{-3}\longrightarrow0.
\end{equation}
The Green contour and exterior Dirichlet energy satisfy the same bounds, while $T_0(-y)=O(y^{-4})$ makes the vacuum boundary term vanish. Thus, the argument does not require a uniform dipole expansion for Lieb--Liniger.
With the physical assumptions of Sec.~\ref{sec:macroscopic-limit}, Eq.~\eqref{eq:app-ll-action-justified} and Appendix~\ref{app:kernel-action} then give 
\begin{equation}
 \mathcal I_{\rm EFP}=(qR)^2\widehat f(r),\qquad
 \widehat f(r)=\frac12\int_{-1}^1t g(t)\,dt,\qquad
 g(t)=\mathfrak g(qt),\quad r=c/q.
\end{equation}

\subsection{Hydrodynamic cutoff and microscopic corrections}
The vanishing contour proves that no finite endpoint term remains in the dispersionless action identity. Keeping a finite cutoff instead estimates the hydrodynamic action omitted near the endpoint.

Set $\lambda=L/q\gg1$ at fixed background. For solutions with a differentiated dipole expansion, the exterior Dirichlet energy and the contour correction are both of relative order $O(\lambda^{-d-1})$. The vacuum area tail is smaller, of order $O(\lambda^{-2d-3})$. The finite-domain identity therefore gives
\begin{equation}\label{eq:app-endpoint-cutoff-tail}
 \frac{\mathcal I_{\rm tail}^{\rm hydro}(\lambda)}{\mathcal I_{\rm hydro}}
 =O(\lambda^{-d-1}).
\end{equation}
The general barrier instead gives $O(\lambda^{d-2a-1})+O(\lambda^{-d-1})$. For Lieb--Liniger, this becomes $O(\lambda^{-3+2\varepsilon})$ for any fixed $0<\varepsilon<3/2$. The correction vanishes as the cutoff approaches the endpoint, although this estimate does not establish the sharper dipole rate $O(\lambda^{-3})$.

For the dipole estimates, the reconstruction $x=R-\mathcal W_y-yT$ gives $|x-R|/R=O(\lambda^{-d-1})$ and $q|T|/R=O(\lambda^{-d-2})$ on the outer contour. These bounds have different powers and do not provide a uniform relation between a spatial cutoff and $\lambda$. Taking the background spacing $\rho_0^{-1}$ as a tentative microscopic scale gives $|x/R-1|\sim(\rho_0R)^{-1}$. Matching this to the leading nonzero spatial displacement suggests a relative hydrodynamic tail of order $(\rho_0R)^{-1}$, corresponding to an $O(R)$ correction to the $O(R^2)$ action at fixed background. 

This estimate concerns only spatial matching. The scale at which hydrodynamics breaks down may depend on the local density and gradient terms, and determining it requires matching in both space and imaginary time. Thus, the power law of $K$ controls the hydrodynamic tail under the assumptions above, but does not determine the microscopic correction or establish global admissibility and minimality. 

\section[Elliptic parametrization of Kazakov, Kostov, and Nekrasov]{%
  Elliptic parametrization of\\
  Kazakov, Kostov, and Nekrasov}\label{app:hoppe-elliptic}
We give an elliptic solution of Eq.~\eqref{eq:main-ll-response},
which we repeat for convenience:
\begin{equation*}
    \int_{-1}^{1}g(s)
    \log\left(1+\frac{r^2}{(t-s)^2}\right)ds=4t,
    \qquad -1<t<1.
\end{equation*}
Here $r=c/q>0$, and we seek an odd function $g$ with integrable
inverse-square-root behavior at $t=\pm1$. Equivalently, the Cauchy
resolvent of Eq.~\eqref{eq:resolvent-definition} satisfies
Eq.~\eqref{eq:hoppe-ll-shift},
\begin{equation*}
    C_+(k)+C_-(k)-C(k+ic)-C(k-ic)=2k,
    \qquad |k|<q.
\end{equation*}
The formulas below give $\chi$, $g$, and the combination
$C_\Delta(z)=C(z)-C(z+ic)$ needed for the observables. We refer to
KKN~\cite[Sec.~6.1]{kazakov1999dparticles} for the derivation. They use a
conformal map to uniformize a quadrant of the complex rapidity plane with a
horizontal slit, mapping it onto the upper half-plane and obtaining the
elliptic parametrization. Throughout, $c,q,R>0$, and the regularity,
reconstruction, and action assumptions of
Appendices~\ref{app:hodograph-action}--\ref{app:ll-endpoint} apply.

\subsection{Normalization, coupling, and rate}\label{app:hoppe-normalization}
The inputs are $c$, the background rapidity $q$ fixed by the Lieb equation~\eqref{eq:problem-ll-thermodynamics}, and the emptiness half-length $R$. To fix the normalization, write $C(z)=A/z+O(z^{-3})$ and introduce KKN's normalized resolvent $W$:
\begin{equation}\label{eq:hoppe-rescaling}
    W(z)=\frac cA C(cz),\qquad
    a=\frac qc=\frac1r,\qquad t_{\rm H}=\frac A{c^2}.
\end{equation}
For real $u=k/c$, Eq.~\eqref{eq:hoppe-ll-shift} becomes
\begin{equation}\label{eq:hoppe-saddle}
    W_+(u)+W_-(u)-W(u+i)-W(u-i)=\frac{2u}{t_{\rm H}},
    \qquad |u|<a,
\end{equation}
where $W_\pm(u)=W(u\pm i0)$, with the normalization
\begin{equation*}
    W(z)\sim\frac1z\qquad(z\to\infty).
\end{equation*}
This is KKN's Eq.~(6.7), with their squared coupling equal to $t_{\rm H}$. We use their regular single-interval solution. Equation~\eqref{eq:resolvent-action} then gives $A=c^2t_{\rm H}=q^2\widehat f$ and $\mathcal I_{\rm EFP}=R^2A$.

We use the elliptic parameter $m$ (the square of the modulus), with conventions
\begin{equation}\label{eq:hoppe-elliptic-conventions}
    F(\phi|m)=\int_0^\phi\frac{d\theta}{\sqrt{1-m\sin^2\theta}},
    \qquad
    E(\phi|m)=\int_0^\phi\sqrt{1-m\sin^2\theta}\,d\theta,
\end{equation}
with $\mathbf K(m)=F(\pi/2|m)$ and $\mathbf E(m)=E(\pi/2|m)$. For $0<m<1$, set
\begin{equation}\label{eq:ll-elliptic-parameters}
    \vartheta=\frac{\mathbf E(m)}{\mathbf K(m)},\qquad
    h=1-\vartheta,\qquad
    \phi=\arcsin\sqrt{\frac hm},\qquad 0<\phi<\frac\pi2.
\end{equation}
Here $0<h<m$, and all roots are positive. KKN's coupling and the support endpoint are
\begin{align}
    t_{\rm H}(m)&=\frac{\mathbf K(m)^4}{3\pi^4}
    \left[-3\vartheta^2+2(2-m)\vartheta-(1-m)\right],
    \label{eq:hoppe-elliptic-coupling}\\
    a(m)&=\frac{\mathbf K(m)E(\phi|m)-\mathbf E(m)F(\phi|m)}{\pi}.
    \label{eq:hoppe-elliptic-endpoint}
\end{align}
The first formula is the second expression in KKN's Eq.~(6.24); the second evaluates their conformal map at the support endpoint.\footnote{In arXiv v2 of Ref.~\cite{kazakov1999dparticles}, the first expression in Eq.~(6.24), written in terms of $\lambda_2$, has a factor-of-four misprint: its prefactor should be $\mathbf K^4/(12\pi^4)$. We use the second expression, consistent with their normalization equation (6.18).}
The function $a(m)$ increases from zero to infinity on $0<m<1$. For the prescribed $r=c/q$, solve $a(m)=q/c$ by a one-dimensional root search. Keep this $m$ fixed in all subsequent formulas. The rate is then
\begin{align}
    r(m)&=\frac1{a(m)},\label{eq:ll-elliptic-r}\\
    \widehat f(r(m))&=r(m)^2t_{\rm H}(m).
    \label{eq:ll-elliptic-rate}
\end{align}
As a normalization check, $a\sim m/8$ and $t_{\rm H}\sim m^2/128$ as $m\to0$, giving $\widehat f\to1/2$. As $m\to1$, $a\sim\mathbf K/\pi$ and $t_{\rm H}\sim2\mathbf K^3/(3\pi^4)$, giving $\widehat f\sim2/(3\pi r)$.

\subsection{The functions \texorpdfstring{$\chi$, $g$, and $C_\Delta$}{chi, g, and C Delta}}\label{app:hoppe-g}
\paragraph*{Real functions.}
To evaluate $\chi$ and $g$, define
\begin{equation}\label{eq:hoppe-real-cut-map}
    U(\theta)=\frac{\mathbf K}{\pi}
    [E(\theta|m)-\vartheta F(\theta|m)].
\end{equation}
It increases from zero to $a$ on $[0,\phi]$, then decreases to zero on $[\phi,\pi/2]$. For each $0<\xi=k/q<1$, invert these two monotone branches:
\begin{equation}\label{eq:hoppe-paired-amplitudes}
    U(\theta_-)=U(\theta_+)=a\xi,
    \qquad 0<\theta_-<\phi<\theta_+<\frac\pi2.
\end{equation}
The auxiliary density is
\begin{equation}\label{eq:hoppe-elliptic-chi}
    \frac{\chi(q\xi)}{Rq}
    =\frac{\mathbf K^2m}{2\pi^2a}
    \left(\sin^2\theta_+-\sin^2\theta_-\right).
\end{equation}
Differentiating $\chi$ with respect to its rapidity argument gives $g(\xi)=-2\chi'(q\xi)/(\pi R)$:
\begin{align}\label{eq:hoppe-elliptic-g}
    g(\xi)&=\frac{2\mathbf K m}{\pi^2}
       [J(\theta_-)-J(\theta_+)], \nonumber \\
    J(\theta)&=\frac{\sin\theta\cos\theta\sqrt{1-m\sin^2\theta}}
       {h-m\sin^2\theta}.
\end{align}
Extend $\chi$ evenly to $[-q,q]$ and $g$ oddly to $[-1,1]$, and set $\chi=0$ outside its support. At $\xi=0$, take $\theta_-=0$, $\theta_+=\pi/2$, and $g(0)=0$. As $\xi\to1^-$, both amplitudes approach $\phi$: $\chi$ vanishes as a square root, while $g$ diverges as an integrable inverse square root. Thus $g$ is evaluated inside the interval, with endpoint behavior understood as a limit, as required in Appendix~\ref{app:resolvent-equivalence}.

In the free-fermion limit, $\chi(q\xi)/(Rq)\to\sqrt{1-\xi^2}$ and $g(\xi)\to2\xi/[\pi\sqrt{1-\xi^2}]$. In the weak-coupling limit, at fixed $0<\xi<1$, $\chi(q\xi)/(Rq)\sim(1-\xi^2)/(2r)$ and $g(\xi)\sim2\xi/(\pi r)$. The latter limits are not uniform at the endpoints: for every finite $r>0$, the square-root endpoint behavior remains.

\paragraph*{Shifted resolvent.}
The same function $U$ gives $C_\Delta$ directly when continued to complex $\theta$:
\begin{align}\label{eq:hoppe-elliptic-Cdelta}
    z&=cU(\theta),\qquad \operatorname{Im}z>0, \nonumber \\
    C_\Delta(z)&=ic\left\{
       \frac{\mathbf K^2}{\pi^2}
       [1-2\vartheta-m\cos^2\theta]
       -\left[U(\theta)+\frac i2\right]^2\right\}.
\end{align}
For a target $z$ in the upper half-plane, first solve $U(i\beta)=i\operatorname{Im}z/c$ for the unique $\beta>0$. Starting at $\theta=i\beta$, follow the root of $U(\theta)=z/c$ continuously along the horizontal segment from $i\operatorname{Im}z$ to $z$. Continue the elliptic integrals from $\theta=0$ along this same branch. This prescription selects the physical solution, for which $C_\Delta(z)\sim icA/z^2$ at infinity. Values on the real axis are approached from above. Equation~\eqref{eq:hoppe-elliptic-Cdelta} supplies $C_\Delta$ directly for the vacuum formula~\eqref{eq:general-boundary-b} and the liquid-profile quadrature~\eqref{eq:ll-profile-quadrature}.

\paragraph*{Cauchy resolvent.}
If $C$ itself is needed, it follows from the Cauchy transform of $\chi$, or equivalently by integration by parts from $g$:
\begin{equation}\label{eq:hoppe-C-quadrature}
    C(z)=\frac q2\int_0^1 g(t)
       \log\frac{z+qt}{z-qt}\,dt,
       \qquad z\notin[-q,q].
\end{equation}
For each $t$, the logarithm is analytic off $[-qt,qt]$ and tends to zero at infinity. This recovers $C(z)\sim q^2\widehat f(r)/z$. The direct elliptic parametrization above is for $C_\Delta$; no separate evaluation of this quadrature is needed for the hydrodynamic reconstruction.

\subsection{Vacuum boundary}\label{app:hoppe-vacuum}
The vacuum boundary requires only real elliptic integrals. With the same $m$ as above, put $\theta=i\operatorname{arsinh}\sqrt{w/m}$ in Eq.~\eqref{eq:hoppe-elliptic-Cdelta} and vary $w\geq0$. Then $U(\theta)=iY(w)$, where
\begin{equation}\label{eq:hoppe-vacuum-Y}
    \begin{aligned}
    Y(w)&=\frac{\mathbf K}{\pi}
    \left[\sqrt{\frac{w(w+1)}{w+m}}
        -E(\alpha|1-m)+hF(\alpha|1-m)\right],\\
    \alpha&=\arctan\sqrt{\frac wm},\qquad
    0\leq\alpha<\frac\pi2,
    \end{aligned}
\end{equation}
and
\begin{equation}\label{eq:hoppe-vacuum-D}
    D(w)=\frac{\sqrt{w(w+m)(w+1)}}{w+h},\qquad
    Y'(w)=\frac{\mathbf K}{2\pi D(w)}.
\end{equation}
Use positive square roots for $w>0$ and the real branch $0\leq\alpha<\pi/2$. The velocity is $v=-cY(w)$, with $Y$ increasing from zero to infinity. In terms of $\eta=|v|/q=Y/a$, the characteristic intercept is $b(\eta)=-2Y+2\mathbf K D/\pi$. The envelope relations~\eqref{eq:main-ll-boundary} give the upper-right vacuum arc ($x,\tau\geq0$):
\begin{align}\label{eq:hoppe-elliptic-vacuum}
    \frac{c\tau(w)}R&=4D(w)D'(w)-2, \nonumber \\
    \frac{x(w)}R&=\frac{2\mathbf K}{\pi}D(w)-4Y(w)D(w)D'(w).
\end{align}
Primes here denote derivatives in $w$. Evaluate $DD'=\tfrac12(D^2)'$; at $w=0$, use $Y=D=0$ and $DD'=m/(2h^2)$. As $w$ increases to infinity, the arc runs from $(0,\tau_c)$ to $(R,0)$, with $x$ increasing and $\tau$ decreasing; $\tau_c$ is given by Eq.~\eqref{eq:ll-elliptic-closing}. Multiply the dimensionless coordinates by $R$ for $x$ and $R/c$ for $\tau$. Reflections give the other quadrants.

The limiting curves follow from the intercepts at fixed finite $\eta$:
\begin{equation*}
    \begin{aligned}
    b(\eta)&\longrightarrow\frac{\eta}{\sqrt{1+\eta^2}},
        &&r\to\infty,\\
    b(\eta)&\longrightarrow\frac2\pi
        \left[\arctan\eta+\frac{\eta}{1+\eta^2}\right],
        &&r\to0.
    \end{aligned}
\end{equation*}
The first gives the free-fermion astroid, with $q\tau_c/R\to1$; this limit is not uniform near $(\pm R,0)$. The second gives the weak-Bose curve of Sec.~\ref{sec:polytropic}, with $q\tau_c/R\to4/\pi$. To obtain the full liquid profile, use $C_\Delta$ from Eq.~\eqref{eq:hoppe-elliptic-Cdelta} in the quadrature~\eqref{eq:ll-profile-quadrature}, together with the Lieb--Liniger distributions $\rho_p(k;Q)$ at varying local $Q$. The background $q$ and elliptic parameter $m$ stay fixed. The quadrature is derived in Appendix~\ref{app:hoppe-profile-quadrature}.

\section{Numerical details}
\label{app:hydro-numerics}
This appendix describes the numerical implementation and convergence checks for the spacetime minimization of Sec.~\ref{sec:hydro-numerics}. The code and numerical data are available at the GitHub repository~\cite{AbanovKhanikati2026code}. Unless stated otherwise, $R=\rho_0=m=\hbar=1$.

\subsection{Constrained variational problem}
Let $s=\widehat c_{s0}\tau$, where $\widehat c_{s0}$ is the sound speed computed from the numerical equation of state, and introduce a displacement $\mathcal U$ by
\begin{equation}
    \rho=\mathcal U_x,
    \qquad
    j=-\widehat c_{s0}\mathcal U_s,
\end{equation}
so continuity holds identically. On the quarter-domain $\Omega_L=[0,L]^2$, the hydrodynamic action is
\begin{equation}\label{eq:num-functional}
\mathcal I_{\rm in}
=4\widehat c_{s0}\int_{\Omega_L}
\left[
\frac{\mathcal U_s^2}{2\mathcal U_x}
+V(\mathcal U_x)
\right]dx\,ds,
\qquad
V(\rho)=
\frac{\epsilon(\rho)-\epsilon(1)-\mu(1)(\rho-1)}
{\widehat c_{s0}^{\,2}}.
\end{equation}
We further impose $\mathcal U_x\geq0$. At zero density, the kinetic term is defined to be zero for vanishing current and infinite otherwise. This allows the vacuum region to emerge naturally during minimization, without imposing its boundary.

The symmetry and emptiness conditions are
\begin{equation}
    \mathcal U(0,s)=0,
    \qquad
    \mathcal U(x,0)=0\quad(0\leq x\leq1).
\end{equation}
On the bottom boundary outside the slit, the zero-current condition follows from the variational principle. 
The factor of $4$ accounts for reflection across both axes. 

\subsection{Outer boundary and exterior action}
At the top and right edges, we impose the leading far-field displacement
\begin{equation}
    \mathcal U_\infty(x,s) = x - \frac{2\alpha x}{x^2+(c_{s0}/\widehat c_{s0})^2s^2}.
\end{equation}
Here $c_{s0}$ is the physical background sound speed and $\widehat c_{s0}$ is its numerical value obtained from the equation of state. The amplitude $\alpha$ is determined by the hydrodynamic relation~\cite[Sec.~5, Eq.~(5.4)]{abanov-hydro}
\begin{equation}\label{eq:num-farfield-amplitude}
    \alpha=\frac{\rho_0}{4\pi c_{s0}}
    \left.\frac{\partial\mathcal I_{\rm EFP}}
    {\partial\rho_0}\right|_{c,R}.
\end{equation}
The derivative is taken at fixed coupling $c$ and slit size $R$. We obtain $\alpha$ from the analytic response, with Eq.~\eqref{eq:num-farfield-amplitude} fixing its normalization. The far-field condition is imposed at the edges of a box of size $L = 13R$, while the vacuum boundary and closing time are obtained from the numerical minimization. 

We evaluate the action from the optimized field and add the leading contribution from outside the box. When the numerical and physical sound speeds agree, this gives
\begin{equation}
    \mathcal I_{\rm out}^{(2)}
    =\frac{(\pi+2)\alpha^2c_{s0}}{\rho_0L^2},
    \qquad
    \mathcal I_{\rm corr}=\mathcal I_{\rm in}+\mathcal I_{\rm out}^{(2)},
\end{equation}
which corrects the leading omitted exterior action.

\subsection{Numerical Lieb--Liniger equation of state}
Although the microscopic coupling $c$ is fixed, the local coupling $\gamma=c/\rho$ varies across the instanton. We compute the equation of state once at $c=1$ and use the scaling relations
\begin{equation}
    \epsilon_c(\rho)=c^3E(\rho/c),
    \qquad
    \mu_c(\rho)=c^2M(\rho/c),
    \qquad
    \mu_c'(\rho)=cM'(\rho/c).
\end{equation}
We interpolate the chemical potential $M(n)$ using cubic Hermite polynomials with derivatives $M'(n)=c_s^2/n$. Integrating and differentiating this interpolant gives the energy and compressibility, ensuring thermodynamic consistency.

Beyond the tabulated range, we use the weak-coupling energy expansion through order $(c/\rho)^{5/2}$~\cite{Tracy2016,Prolhac2017}, smoothly matched to the numerical equation of state. Independent Bethe quadratures give relative discrepancies below $10^{-7}$. The Tonks--Girardeau limit is treated exactly, with $\epsilon_{\rm TG}=\pi^2\rho^3/6$.

\subsection{Discretization and optimization}
We represent $\mathcal U$ using quadratic B-splines, enforcing nonnegative density and allowing reduced smoothness at the slit endpoint. Near the slit, we impose $\mathcal U=O(s^2)$ to keep the kinetic action finite.

We start with 64 graded intervals per axis, concentrated near $x = 0, 1$ and $s = 0$, and extend the mesh to $L = 13$ with 32 larger intervals. We refine adaptively near the vacuum boundary and its cusps, with additional refinement near the slit endpoint to resolve the density rise from $0.02\rho_0$ to $0.8\rho_0$ with at least 12 intervals. The final meshes contain $(5.07$--$5.13)\times10^5$ coefficients, depending on the coupling. All refinements use only the preceding numerical solution, without input from the analytic arctic curve.

After fixing the boundary coefficients, we collect the remaining spline coefficients in $\boldsymbol a$. Their positivity constraints are written as $\boldsymbol z=C\boldsymbol a+\boldsymbol d\geq0$ and enforced using a logarithmic barrier,
\begin{equation}
    F_{h,\nu}=F_h-\nu\sum_\ell w_\ell^{(c)}\log z_\ell,
\end{equation}
where $F_h$ is the discrete action. We minimize this using sparse Newton steps with a line search that preserves positivity. On the final meshes, the discrete optimality gap is bounded by $3\times10^{-6}$ of the action.

We decrease the barrier parameter from $10^{-3}$ to $10^{-9}$, repeating the final steps on each refined mesh. The optimization uses four-point Gauss--Legendre quadrature, while the final action is evaluated with eight-point quadrature and the exterior correction. Since the barrier leaves a small positive density in the vacuum region, numerical density plots and contours are interpreted accordingly.

\subsection{Extraction of the vacuum boundary}
Finite density thresholds and mesh resolution make fixed-density contours unreliable for locating the vacuum boundary. Instead, we extrapolate from the liquid side using the near-boundary expansion
\begin{equation}
    x=x_e+\mu(\rho)T_0'(v_e)+\cdots.
\end{equation}
At each time, we locate the first spatial knot with $\rho\geq0.02\rho_0$ and consider three fitting windows, spanning knots 3--8, 4--10, and 5--12 beyond it. We fit $x$ as a linear and quadratic function of $\mu$, with the intercept at $\mu=0$ giving an estimate of $x_e$.

We reject fits with nonmonotone density, density exceeding $0.8\rho_0$, or excessive extrapolation. A boundary point is accepted only if all three linear fits pass these checks, and we take the median of their intercepts. The quadratic fits provide a check on sensitivity. Unresolved regions near the cusps are omitted.

We determine the closing time separately using the near-vacuum expansion of the hodograph time, 
\begin{equation}
    T(\rho,v)=T_0(v)-F_1(\rho)T_0''(v) + F_2(\rho)T_0^{(4)}(v)-\cdots,
\end{equation}
where $F_1$ and $F_2$ are obtained recursively from the numerical equation of state. We fit over several density ranges and at two truncation orders, taking the median of the admissible low-density estimates. 

\subsection{Convergence and limitations}
The numerical calculations were performed at
\begin{equation}
    \gamma_0=\infty,\;100,\;10,\;1,\;0.01.
\end{equation}
Table~\ref{tab:hydro-convergence} summarizes the mesh refinement and comparison errors for each coupling. 

\begin{table}[htbp]
\centering
\small
\begin{tabular}{@{}rrrrrrr@{}}
\toprule
$\gamma_0$ & $\delta\mathcal I$ (\%) & $e_{\mathcal I}$ (\%) & $\delta\tau_c$ (\%) & $10^3 e_{x,\rm RMS}$ & $10^3 e_{x,\max}$ & Coverage (\%)\\
\midrule
$\infty$ & 0.0005 & 0.0226 & 0.0003 & 0.755 & 2.104 & 97.9 \\
$100$ & 0.0004 & 0.0189 & 0.0002 & 0.757 & 2.098 & 97.9 \\
$10$ & 0.0001 & 0.0136 & $<0.0001$ & 0.797 & 2.183 & 97.9 \\
$1$ & 0.0001 & 0.0076 & $<0.0001$ & 0.952 & 2.600 & 96.6 \\
$0.01$ & 0.0001 & 0.0072 & 0.0001 & 1.280 & 3.948 & 86.8 \\
\bottomrule
\end{tabular}
\caption{Numerical convergence and boundary accuracy. The finest meshes contain $(5.07$--$5.13)\times10^5$ coefficients, compared with $(3.79$--$3.85)\times10^5$ on the preceding meshes. Both include targeted refinement near the slit endpoint. The quantities $\delta\mathcal I$ and $\delta\tau_c$ denote absolute relative changes between successive meshes, while $e_{\mathcal I}=\mathcal I_{\rm corr}/\mathcal I_{\rm analytic}-1$ is the relative action error. Boundary errors are measured as $|x_{\rm num}-x_{\rm analytic}|/R$ at matching physical times. Coverage denotes the fraction of the time interval with accepted boundary points.}
\label{tab:hydro-convergence}
\end{table}

Since the final refinements focus on the slit endpoint, the last-mesh changes $\delta\mathcal I$ and $\delta\tau_c$ in Table~\ref{tab:hydro-convergence} measure sensitivity to endpoint resolution. Across these refinements, the fitted closing time changes by only $0.006$--$0.008\%$. On the accepted boundary points, the RMS deviation from the analytic curve is of order $10^{-3}R$, with a maximum roughly three times larger near the closing point. About $2$--$13\%$ of the time interval remains unresolved and is excluded from these error estimates.

The numerical action decreases with each refinement. The final corrected action exceeds the analytic value by $0.007$--$0.023\%$. The changes become progressively smaller with endpoint refinement, indicating that the remaining discrepancy is not dominated by endpoint resolution. For $\gamma_0\leq10$, the last change is below the relative optimization gap bound of about $3\times10^{-6}$, while quadrature differences are of order $10^{-8}$. However, the action has not reached a mesh-independent plateau, and no continuum extrapolation is performed.

The largest remaining boundary errors come from limited resolution near the closing point; the source of the remaining action excess has not been isolated. The numerical results test the dispersionless variational problem but do not establish global hodograph invertibility or control microscopic dispersive corrections.

\section{Polytropic solution and singular-kernel checks}\label{app:polytropic-details}
We derive the vacuum curve of Sec.~\ref{sec:polytropic} and show that the singular-kernel representation reproduces the same potential and action. The direct solution is valid for all $d>0$, while the singular-kernel representation applies to $d=2n+1$, with real $n\geq0$ and $a>0$, corresponding to $1<\gpol\leq3$.

\subsection{Vacuum boundary}
The regular Bessel modes extend the boundary values $\mathcal W(Q,0)=-F(Q)$ of Eq.~\eqref{eq:polytropic-FQ} into the hodograph quadrant. At $Q=0$, this radial Poisson  extension gives the potential along the vacuum boundary in the upper-right quadrant of physical spacetime, 
\begin{equation}\label{eq:polytropic-wvac}
    \mathcal W_{\mathrm{vac}}(v) = \frac{2R}{\pi}\int_0^q \frac{Q^{d-1}v \sqrt{q^2 - Q^2}}{(Q^2 + v^2 )^{(d+1)/2}}\,dQ.
\end{equation}
For $d>0$, the integral converges at $Q=0$ and can be differentiated at $v<0$. Defining the characteristic intercept $X(v)=R+\partial_v\mathcal W_{\rm vac}(v)$, differentiation and integration by parts give
\begin{equation}
 X'(v)=-\frac{R}{q}N_d\left(1+\frac{v^2}{q^2}\right)^{-(d+2)/2},
\end{equation}
where $N_d$ is defined in Sec.~\ref{sec:polytropic}. Integrating from $X(-\infty)=R$ gives $X(v)=Rb_d(|v|/q)$, since
\begin{equation*}
 N_d\int_0^\infty(1+s^2)^{-(d+2)/2}\,ds=1.
\end{equation*}
The envelope relations $\tau=-X'$ and $x=X-vX'$ give the arctic curve in Sec.~\ref{sec:polytropic}, with $\tau>0$ because $X'<0$.

\subsection{Supported powers and thermodynamics}
To solve the singular dressing and boundary equations, we extend the functions by zero and apply the whole-line Fourier multiplier $|\partial_\lambda|^{2s}$. The supported-power identity is~\cite{dyda2017fractional}
\begin{equation}
 |\partial_\lambda|^{2s}(Q^2-\lambda^2)_+^s
 =\frac{2^{2s}\Gamma(s+1)\Gamma(s+\tfrac12)}{\sqrt\pi}
 =\Gamma(2s+1),\qquad |\lambda|<Q,\quad s\geq0.
\end{equation}
For $s=0$, the supported power is the interval indicator and the operator is the identity. For $s>0$, the first equality follows from the degree-zero case of Theorem~3 in Ref.~\cite{dyda2017fractional}, and the second from the gamma duplication formula. Setting $s=n$ and $s=n+1/2$ gives

\begin{equation}
    |\partial_\lambda|^{2n}(Q^2-\lambda^2)_+^n
    =\Gamma(2n+1),
    \qquad |\lambda|<Q,
\end{equation}
\begin{equation}
    |\partial_\lambda|^{2n+1}(q^2-\lambda^2)_+^{n+1/2}
    =\Gamma(2n+2),
    \qquad |\lambda|<q.
\end{equation}
These identities give the functions in Sec.~\ref{sec:polytropic}: the first solves $(1-\mathcal T_n)\rho_p=1/(2\pi)$, while the second solves the equation for $\chi_n$. In both cases, the functions must be extended by zero when applying the full operators, including their contact terms.

Integrating $\rho_p$ and $\lambda^2\rho_p/2$ over $[-Q,Q]$ gives the density and energy. With the substitution $\lambda=Qt$, both integrals reduce to beta functions: 
\begin{equation}
 \rho(Q)=\frac{Q^{2n+1}}{2^{2n+1}a\Gamma(n+\tfrac12)\Gamma(n+\tfrac32)},
 \qquad \epsilon(Q)=\frac{\rho(Q)Q^2}{2(2n+3)}.
\end{equation}
It follows that $\mu=\epsilon_Q/\rho_Q=Q^2/[2(2n+1)]$ and $c_s^2=\rho\mu_Q/\rho_Q=Q^2/(2n+1)^2$. Thus, $Q$ is the acoustic coordinate, with $c_s=Q/(2n+1)$.

\subsection{Reconstruction and action}
Using the Fourier convention $\widehat\chi_n(p)=\int e^{-ip\lambda}\chi_n(\lambda)\,d\lambda$, we obtain
\begin{equation}
 \widehat\chi_n(p)=\frac{R\sqrt\pi\Gamma(n+\tfrac32)}{a\Gamma(2n+2)}
 \left(\frac{2q}{|p|}\right)^{n+1}J_{n+1}(q|p|).
\end{equation}
Multiplication by $ap^{2n}/\pi$ gives the spectral amplitude. For $p>0$, gamma function identities give
\begin{equation}
 \frac{ap^{2n}\widehat\chi_n(p)}{\pi}
 =\frac{Rq^{n+1}}{2^n\Gamma(n+1)}p^{n-1}J_{n+1}(pq).
\end{equation}
Together with the regular modes $B_p(Q)$ at $d=2n+1$, this is the Bessel expansion of the direct solution with boundary values~\eqref{eq:polytropic-FQ}. It therefore reproduces the same mixed data and the potential on the vacuum boundary~\eqref{eq:polytropic-wvac}, without using the dressed endpoint quotient for a singular kernel.

The action moment is another beta integral:
\begin{equation}
 \frac{R^2}{2}\int_{-q}^q\lambda\mathfrak g_n(\lambda)\,d\lambda
 =\frac{R^2q^{2n+2}}{2^{2n+1}a\Gamma(n+1)\Gamma(n+2)}.
\end{equation}
Eliminating $a$ using the density at $Q=q$ and $c_{s0}=q/(2n+1)$ gives
\begin{equation}
 \frac{\mathcal I}{\rho_0c_{s0}R^2}
 =\frac{2}{n+1}\left[\frac{\Gamma(n+\tfrac32)}{\Gamma(n+1)}\right]^2.
\end{equation}
With $d=2n+1$ and $\Gamma(z+1)=z\Gamma(z)$, this equals the direct hydrodynamic action in Sec.~\ref{sec:polytropic}. The equality verifies the action moment for this singular family, without extending the regular-kernel endpoint argument to arbitrary distributional kernels.

\bibliography{emptiness}

\nolinenumbers
\end{document}